\documentclass[lettersize,journal]{IEEEtran}
\usepackage{amsmath,amsfonts}
\usepackage{algorithmic}
\usepackage{algorithm}
\usepackage{array}
\usepackage[caption=false,font=normalsize,labelfont=sf,textfont=sf]{subfig}
\usepackage{textcomp}
\usepackage{stfloats}
\usepackage{url}
\usepackage{verbatim}
\usepackage{graphicx}
\usepackage{cite}
\usepackage{multirow}
\usepackage{booktabs}%提供命令\toprule、\midrule、
\usepackage[colorlinks,
linkcolor=blue,
urlcolor = blue,
citecolor=blue
]{hyperref}

\begin{document}
\begin{sloppypar}
\title{Semi-Supervised Learning for AVO Inversion with Strong Spatial Feature Constraints}
% \title{AVO Inversion with Multi-Input Multi-Output and strong spatial feature constraints}
\author{Yingtian Liu, Mingwei Wang, Junheng Peng, and Yong Li
\thanks{Yingtian Liu, Mingwei Wang, and Junheng Peng are with the College of Geophysics, Chengdu University of Technology, Chengdu 610059, China (e-mail: yingtianliu06@outlook.com; WangMingwei023478@outlook.com; 351587229@qq.com). Yong Li is with the State Key Laboratory of Oil and Gas Reservoir Geology and Exploitation and the College of Geophysics, Chengdu University of Technology, Chengdu 610059, China (e-mail: liyong07@cdut.edu.cn).}}


% \markboth{IEEE TRANSACTIONS ON GEOSCIENCE AND REMOTE SENSING, Vol.  , 2025}%


\maketitle

\begin{abstract}
% As a data-driven approach, neural networks are widely used in prestack amplitude variation with offset (AVO) inversion to address its nonlinearity and ill-posed nature. 
One-dimensional convolution is a widely used deep learning technique in prestack amplitude variation with offset (AVO) inversion; however, it lacks lateral continuity. Although two-dimensional convolution improves lateral continuity, due to the sparsity of well-log data, the model only learns weak spatial features and fails to explore the spatial correlations in seismic data fully. To overcome these challenges, we propose a novel AVO inversion method based on semi-supervised learning with strong spatial feature constraints (SSFC-SSL). First, two-dimensional predicted values are obtained through the inversion network, and the predicted values at well locations are sparsely represented using well-log labels. Subsequently, a label-annihilation operator is introduced, enabling the predicted values at non-well locations to learn the spatial features of well locations through the neural network. Ultimately, a two-way strong spatial feature mapping between non-well locations and well locations is achieved. Additionally, to reduce the dependence on well-log labels, we combine the semi-supervised learning strategy with a low-frequency model, further enhancing the robustness of the method. Experimental results on both synthetic example and field data demonstrate that the proposed method significantly improves lateral continuity and inversion accuracy compared to one- and two-dimensional deep learning techniques.
\end{abstract}


\section{Introduction}
\IEEEPARstart{A}{mplitude} variation with offset (AVO) analysis has long been a crucial tool in hydrocarbon exploration and reservoir characterization, providing valuable insights into the elastic properties of subsurface formations \cite{simmons1996waveform},\cite{alemie2011high},\cite{meng2021avo}. Zoeppritz \cite{zoeppritz1919reflection} conducted the first theoretical investigation into AVO technology. By examining the variation in reflection coefficients with incidence angles, he analyzed the parameter changes in the media on either side of the reflection interface and formulated the Zoeppritz equations. Although these equations provide precise calculations of reflection coefficients, their complex expressions make them challenging to apply directly to AVO inversion of real-world data. Consequently, researchers have developed simplified versions of the Zoeppritz equations from various perspectives  \cite{bortfeld1961approximations}, \cite{aid1980quantitative}, \cite{shuey1985simplification}, \cite{smith1987weighted}, \cite{goodway1997improved}, \cite{gray2000application}. 

% \begin{figure*}[!t]
% \centering
% \includegraphics[width=6in]{figure1}
% \caption{Architectures of different DL methods. (a) Inversion method with single-trace input and single-trace output. (b) Inversion method with multi-trace inputs and a single-trace output. (c) Inversion method with multi-trace inputs and multi-trace outputs.}
% \label{fig1}
% \end{figure*}

\begin{figure*}[!t]
\centering
\includegraphics[width=5.2in]{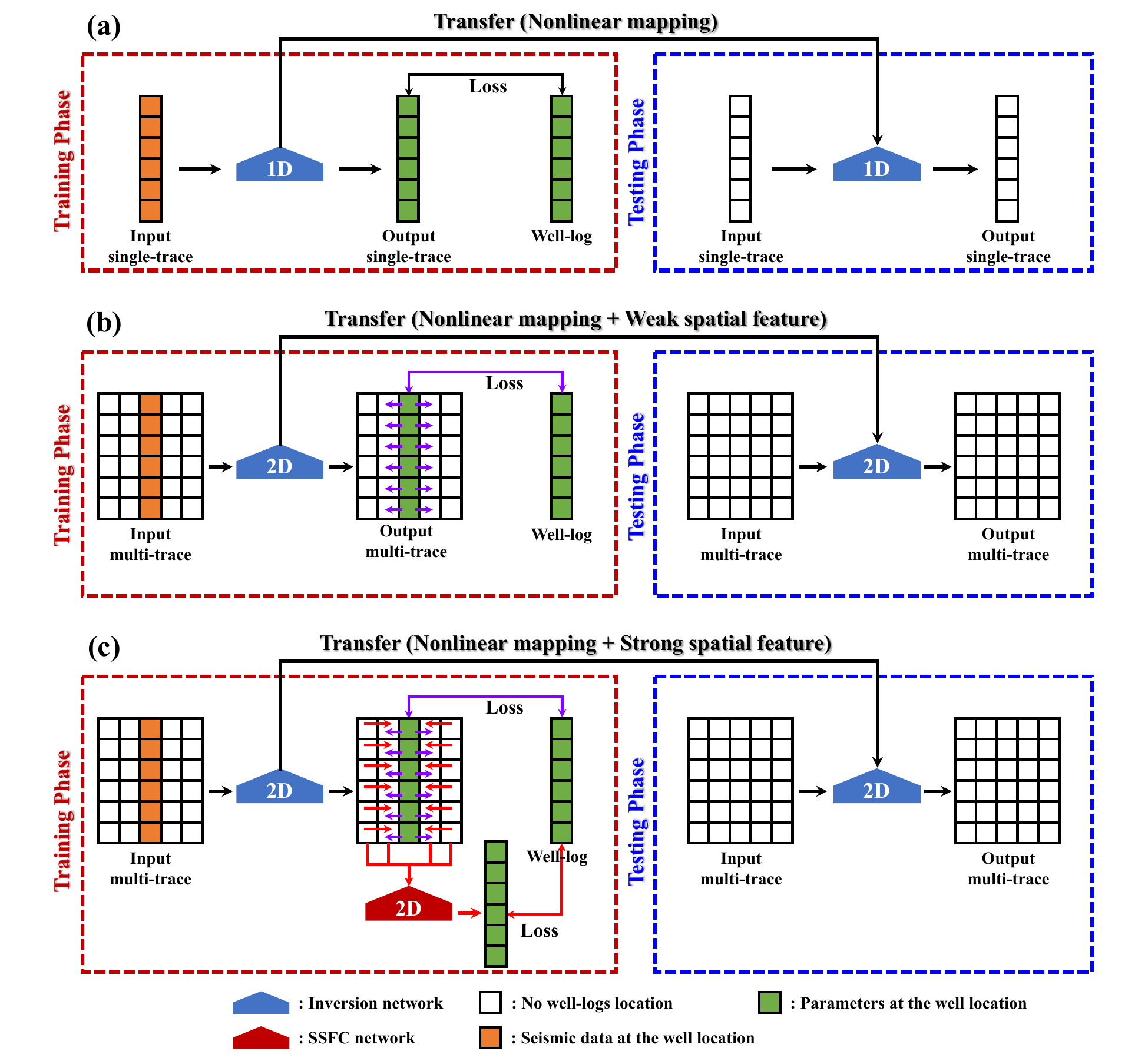}
\caption{Training and testing workflows for seismic inversion methods. (a) 1D data-driven seismic inversion, (b) 2D data-driven seismic inversion, and (c) the SSFC-SSL seismic inversion method.}
\label{train_test}
\end{figure*}

Model-driven AVO inversion is a method to extract seismic parameters such as P-wave velocity, S-wave velocity, and density from prestack seismic data based on AVO theory. 
% Traditional model-driven AVO inversion methods rely on the accurate estimation of background trends and rock physics parameters, such as linearized inversion techniques.
The traditional model-driven AVO inversion method used a linear method to establish the mapping between the reflection coefficient and parameters. Ikelle \cite{ikelle1995linearized} proposed a linear AVO inversion algorithm based on the least-squares method for processing 3D multi-migration seismic data. Downton and Ursenbach \cite{downton2006linearized} proposed a linearized AVO inversion method that can accurately deal with the reflection coefficients of the above critical angles, considering the amplitude and phase changes with the migration.  
% Kuzma et al. \cite{downton2006linearized} proposed a method for approximating inversion relationships using support vector machines (SVM), trained SVM models to simulate the inversion process of nonlinear Zoeppritz equations, and demonstrated the advantages of SVM in capturing nonlinear relationships, reducing computational effort, and improving model estimation accuracy.
However, AVO inversion is an ill-posed problem with many difficulties, such as limited data bandwidth, noise corruption, and incomplete data coverage \cite{yin2016avo}, \cite{she2019avo}. To overcome this problem, many methods have been proposed, including Bayesian probabilistic inversion \cite{zong2016joint}, \cite{li2022hierarchical}, exhaustive searching \cite{jensen2016quantitative},\cite{misra2008global}. Gradient-based algorithms, such as the limited-memory BFGS (L-BFGS) method \cite{ahmed2022constrained},\cite{ahmed2023frequency}, are commonly employed for nonlinear inverse problems. These techniques typically involve the construction of a geological or petrophysical model, which is then used to derive synthetic seismic responses that are matched to the observed data through an iterative optimization process \cite{feng2025acoustic}, \cite{avseth2016avo}, \cite{du2013pp}. While these inversion methods have demonstrated promising results, they are computationally intensive, and their outcomes are highly sensitive to the selection of the initial model \cite{li2022pertinent}, \cite{niu2021data}, \cite{zhang2022deep}.

% Traditional prestack AVO inversion methods generally depend on physical models, such as the Zoeppritz equations and their approximations, like the Aki-Richards approximation (Aki and Richards, 1980). Buland et al. (2003) proposed a linear AVO inversion method based on Bayesian theory that inverts prestack gathers data into P-wave velocity, S-wave velocity, and density. Wang et al. (2009) used a generalized linear algorithm and created a target function based on the exact Zoeppritz equations to invert prestack seismic data into elastic parameters, including P-wave velocity, S-wave velocity, and density. Zhou et al. (2021) proposed a prestack AVO inversion method based on the exact Zoeppritz equations and performed three-parameter inversion tests in a real-world field. Wang et al. (2022) introduced a gradient structural similarity method to calculate the structural similarity between prestack and post-stack inversion results, improving the accuracy of elastic parameter prediction.

Deep learning (DL), known for its ability to represent complex features and perform nonlinear mapping, has become a key research focus in seismic exploration \cite{yang2019deep}, \cite{wang2020current}. In this field, DL techniques have achieved promising results in various aspects, including reservoir parameter prediction \cite{chen2020deep}, \cite{masroor2023multiple}, fault identification \cite{xiong2018seismic}, \cite{wang2024fast}, noise attenuation \cite{liao2023twice}, \cite{zhao2018low}, and stratigraphic interpretation \cite{di2020seismic}, \cite{gu2023semi}.
%1D有监督
In recent years, DL has also been developed in AVO inversion. Das et al. \cite{das2019convolutional} employed a one-dimensional convolutional neural network (CNN) for seismic inversion. Mustafa et al. \cite{mustafa2019estimation} proposed a workflow for impedance prediction utilizing temporal convolutional networks (TCN). Cao et al. \cite{cao2018elastic} proposed a DL-based AVO inversion method with Dropout regularization, enhancing the prediction accuracy and stability of P-wave velocity, S-wave velocity, and density. As shown in Fig. \ref{train_test}(a), the earliest DL inversion method was to extract the characteristics of prediction parameters from single-trace seismic data and utilize well-log data to constrain these characteristics, thereby establishing a non-linear mapping between seismic data and elastic parameters. \cite{zhang2021robust}, \cite{li2019deep}, \cite{zheng2019applications}. 

%1D半监督
In practical applications, the availability of labeled data is limited due to the high cost of well-log data. 
% scholars have incorporated a substantial amount of seismic data to constrain the inverse problem. 
% Transfer learning is introduced into AVO inversion \cite{meng2021avo}, \cite{sun2024seismic}， \cite{liu2024high}. It can facilitate the solution to converge to the true value more easily by taking the low-frequency components as constraint terms.
To solve the challenge of limited labeled data, scholars have introduced various constraints. Alfarraj and AlRegib \cite{alfarraj2019semisupervised} proposed a semi-supervised recurrent neural network (RNN) framework based on seismic forward modeling. Liu et al. \cite{liu2024nash} integrated the semi-supervised temporal convolutional network with Nash game theory \cite{navon2022multi} to effectively address the gradient conflict problem in AVO inversion. Physics-guided neural networks (PGNN) have also been applied to both prestack and post-stack AVO inversion \cite{sun2021physics}, \cite{biswas2019prestack}, \cite{ge2024deep}.  
Many semi-supervised inversion methods incorporating model-based forward modeling have been developed to address various geological inverse problems \cite{adler2019deep}, \cite{guo2019application}, \cite{fabien2020seismic}. 
These methods incorporate seismic data constraints through model-based forward modeling but depend on the precise determination of the wavelet, which is challenging in practical applications. 
Yuan et al. \cite{yuan2022double} proposed employing a network as an alternative to the traditional physical forward modeling process. Shi et al. \cite{shi2024seimic}  introduced a closed-loop seismic inversion method that incorporates a forward network to constrain the inversion network. Wang et al. \cite{wang2023avo} proposed an AVO inversion method based on the closed-loop seismic inversion framework, capable of accurately inverting the P-wave velocity, S-wave velocity, and density. Although these methods effectively establish the mapping between single-trace seismic data and corresponding single-trace parameters, they overlook the spatial correlations among the traces, resulting in significant vertical artifacts in the inversion results.

\begin{figure}[!t]
\centering
\includegraphics[width=3.2in]{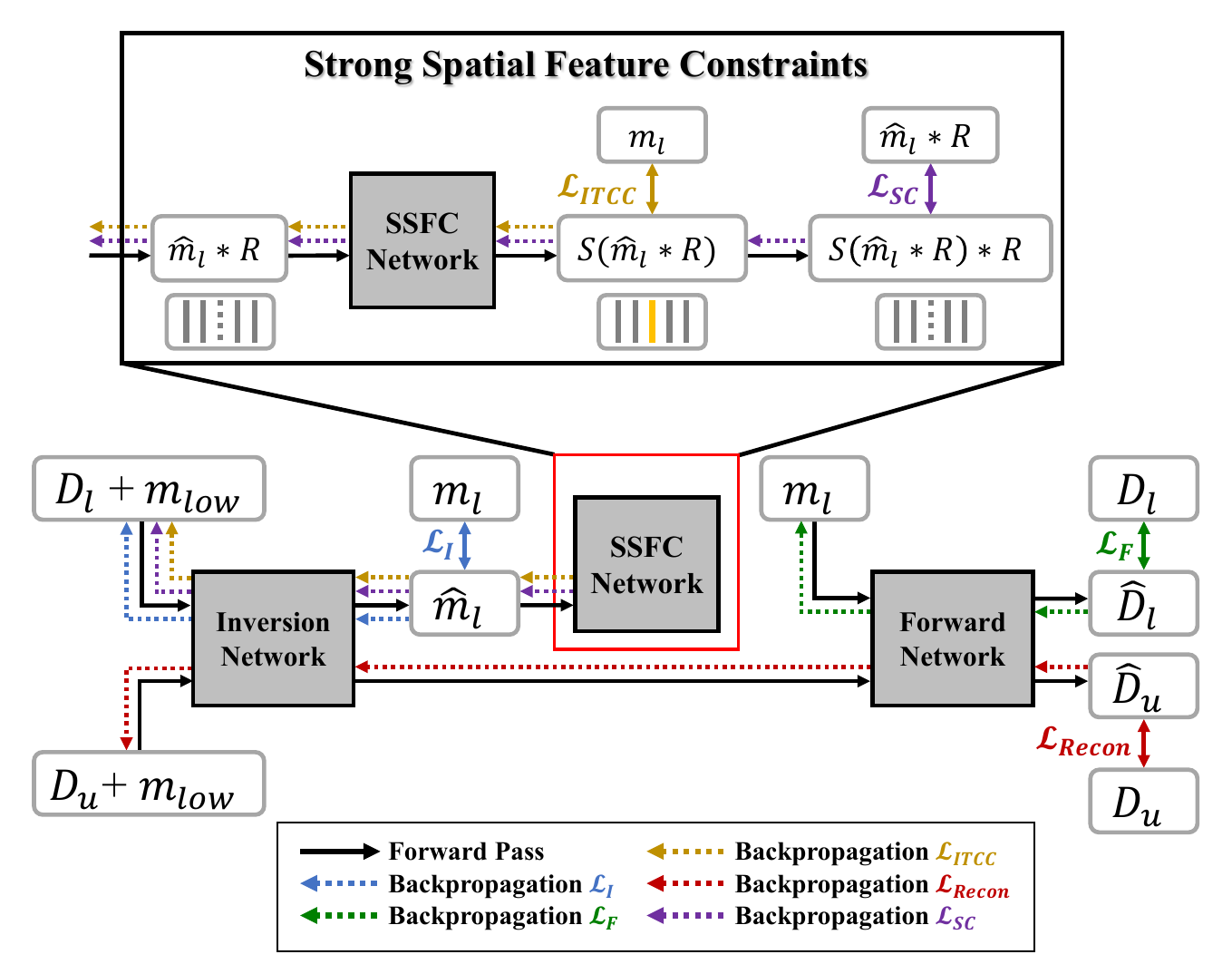}
\caption{Training process of SSFC-SSL.}
\label{fig2}
\end{figure}

%2D有监督
To improve lateral continuity, high-dimensional convolutional kernels were introduced to establish correlations between multi-trace seismic data and multi-trace parameters.
Wu et al. \cite{wu2021deep} proposed an improved DL method based on two-dimensional CNN that significantly improves the stability of impedance prediction.
%2D半监督
Wang et al. \cite{wang2022seismic} introduced a two-dimensional semi-supervised inversion method that improves the accuracy and spatial continuity of seismic inversion.
Mustafa et al. \cite{mustafa2021joint} proposed a joint learning strategy that integrates a two-dimensional temporal convolutional network (TCN) and spatial contexts, improving the robustness and spatial consistency of the estimated parameters. Li et al. \cite{li2024transinver} developed a three-dimensional residual structure, which is used for efficient 3D seismic impedance inversion. These methods process multi-trace seismic data using high-dimensional convolution kernel and output single-trace prediction parameters. As shown in Fig. \ref{fig2}(b), during training, these methods take multi-trace seismic data as input and produce predicted multi-trace parameters as output. The predicted parameters at well locations are constrained by well-log data, In contrast, those at non-well locations are derived from the sparse representation of well-log information and spatial reasoning facilitated by the two-dimensional network. Although these methods improve the lateral continuity of the inversion results to some extent, we have observed that they primarily utilize weak one-way spatial correlations from well to non-well locations. Consequently, they fail to capture the stronger spatial features within the seismic data.

To fully explore the strong spatial features in seismic data and improve the accuracy of prestack seismic AVO inversion, we propose a semi-supervised seismic inversion method based on strong spatial feature constraints (SSFC-SSL). SSFC-SSL combines an inversion network, a forward network, and a strong spatial feature constraint (SSFC) network. 
As shown in Figure \ref{train_test} (c), multi-trace inputs can yield multi-trace outputs through the inversion network. Specifically, the predicted parameters at well locations undergo feature learning through well-log data, while the parameters at non-well locations are inferred laterally by the network. Subsequently, the predicted parameters at non-well locations are used to infer the parameters at well locations via the SSFC network, thereby establishing a bidirectional strong spatial relationship between well and non-well locations. Additionally, considering the limited number of labels in practice, we constructed a forward-modeling network to make full use of the seismic data at non-well locations and improve the generalization ability of the method.
% Additionally, a low-frequency model constraint is incorporated to improve inversion accuracy.

The article is organized as follows. First, we introduce the theory of SSFC-SSL, including its framework, training workflow, and network structure. Then, we compare the AVO inversion results of the L-BFGS method, the one-dimensional deep learning method, the two-dimensional deep learning method, and the proposed method. Next, the field data application is provided to demonstrate the feasibility of our method for field data. Finally, we draw the key conclusions. 

\section{Theory}
\subsection{Prestack AVO inversion and forward}
\begin{figure}[!t]
\centering
\includegraphics[width=2.5in]{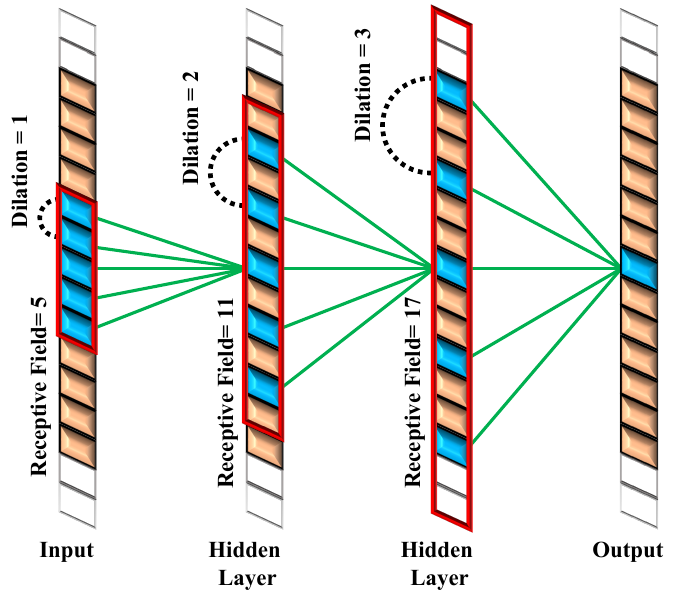}
\caption{The structure of dilated convolution.}
\label{fig3}
\end{figure}
\begin{figure}[!t]
\centering
\includegraphics[width=3in]{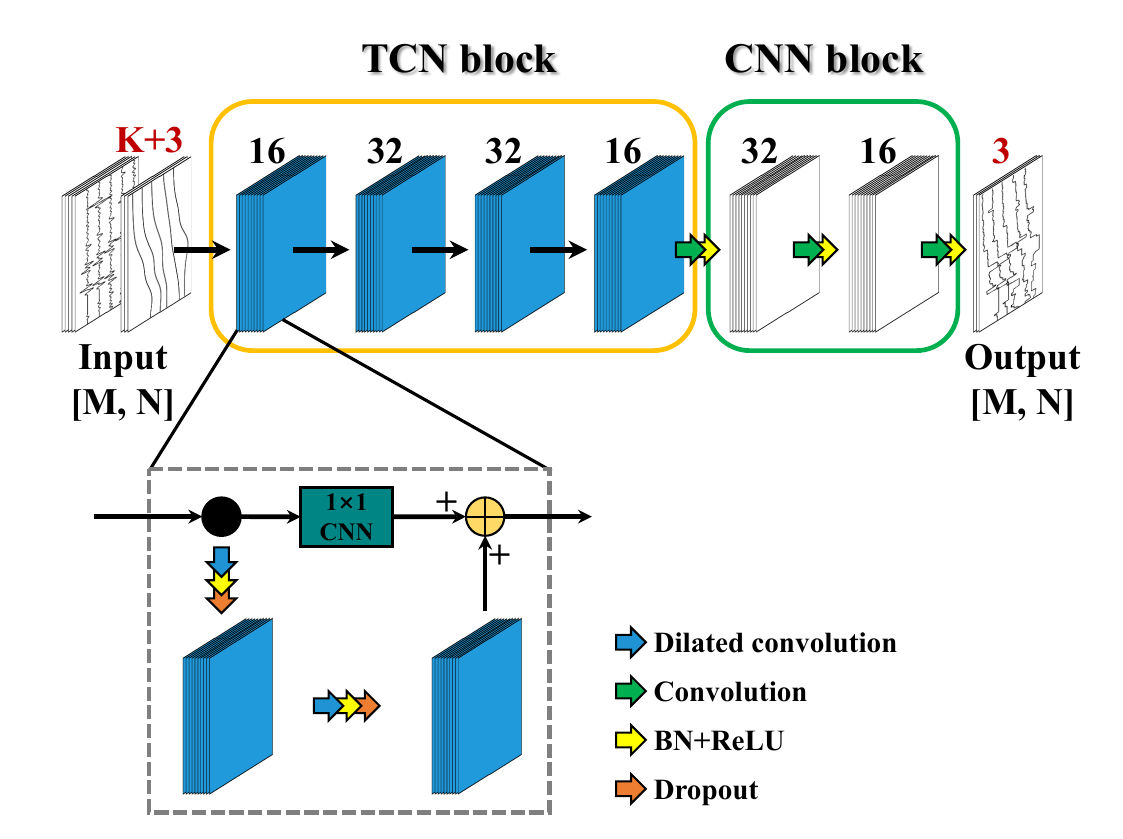}
\caption{Architecture of the inversion network.}
\label{fig4}
\end{figure}
\begin{figure}[!t]
\centering
\includegraphics[width=2in]{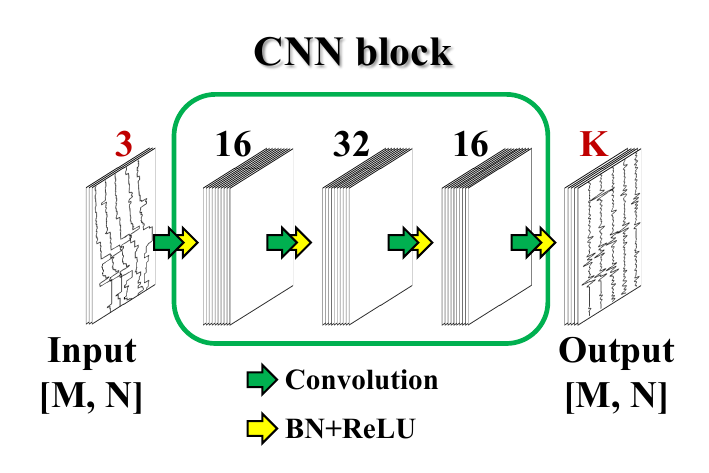}
\caption{Architecture of the forward network.}
\label{fig5}
\end{figure}
\begin{figure}[!t]
\centering
\includegraphics[width=2in]{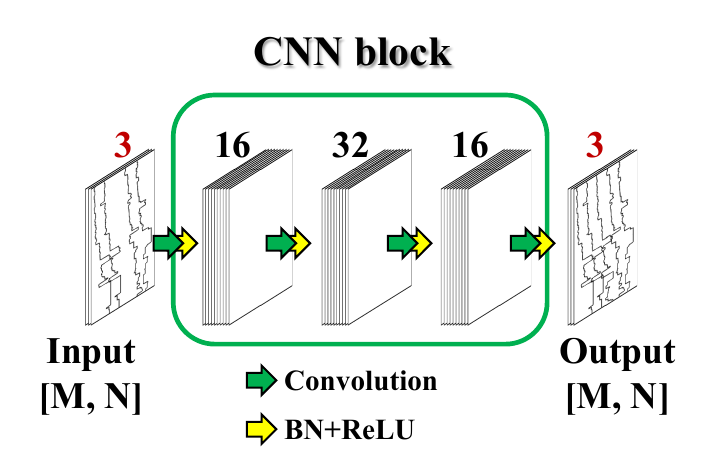}
\caption{Architecture of the SSFC network.}
\label{fig6}
\end{figure}
\begin{figure*}[!t]
\centering
\includegraphics[width=7in]{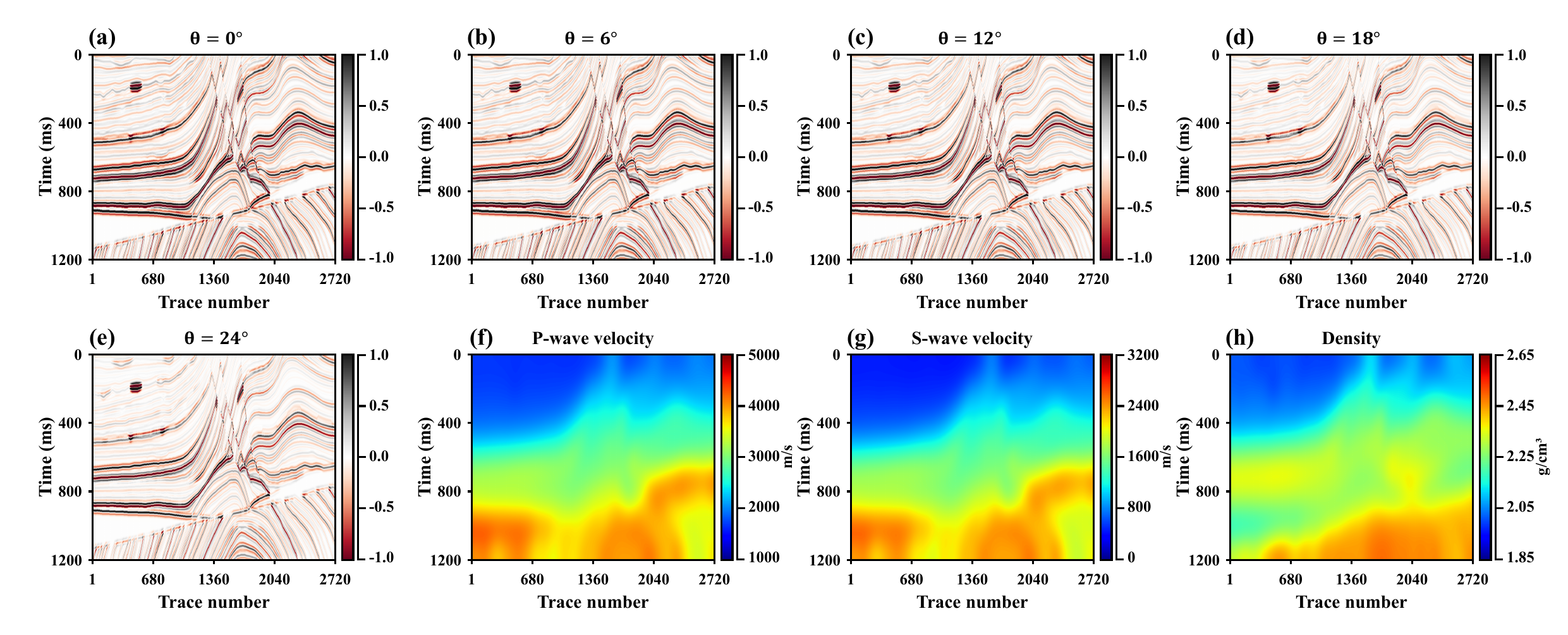}
\caption{Synthetic angle gather profiles with different incident angles of (a) $0^{\circ}$, (b) $6^{\circ}$, (c) $12^{\circ}$, (d) $18^{\circ}$, and (e) $24^{\circ}$. Low-frequency information profile of (f) P-wave velocity, (g) S-wave velocity, and (h) density.}
\label{fig7}
\end{figure*}
\begin{figure*}[!t]
\centering
\includegraphics[width=6.5in]{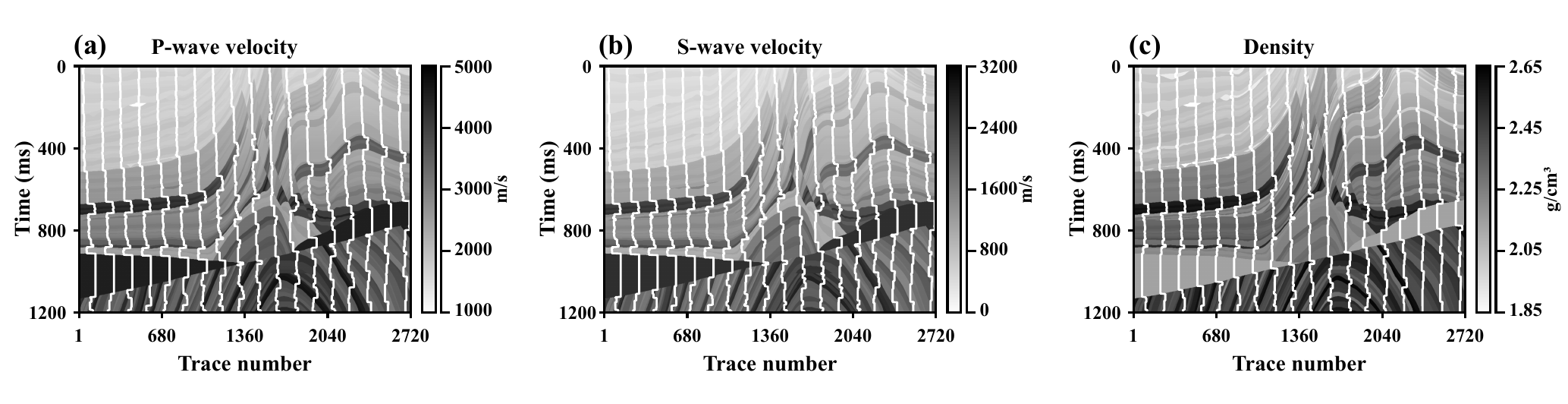}
\caption{Training traces of (a) P-wave velocity, (b) S-wave velocity, and (c) density.}
\label{fig8}
\end{figure*}
\begin{figure}[!t]
\centering
\includegraphics[width=3.5in]{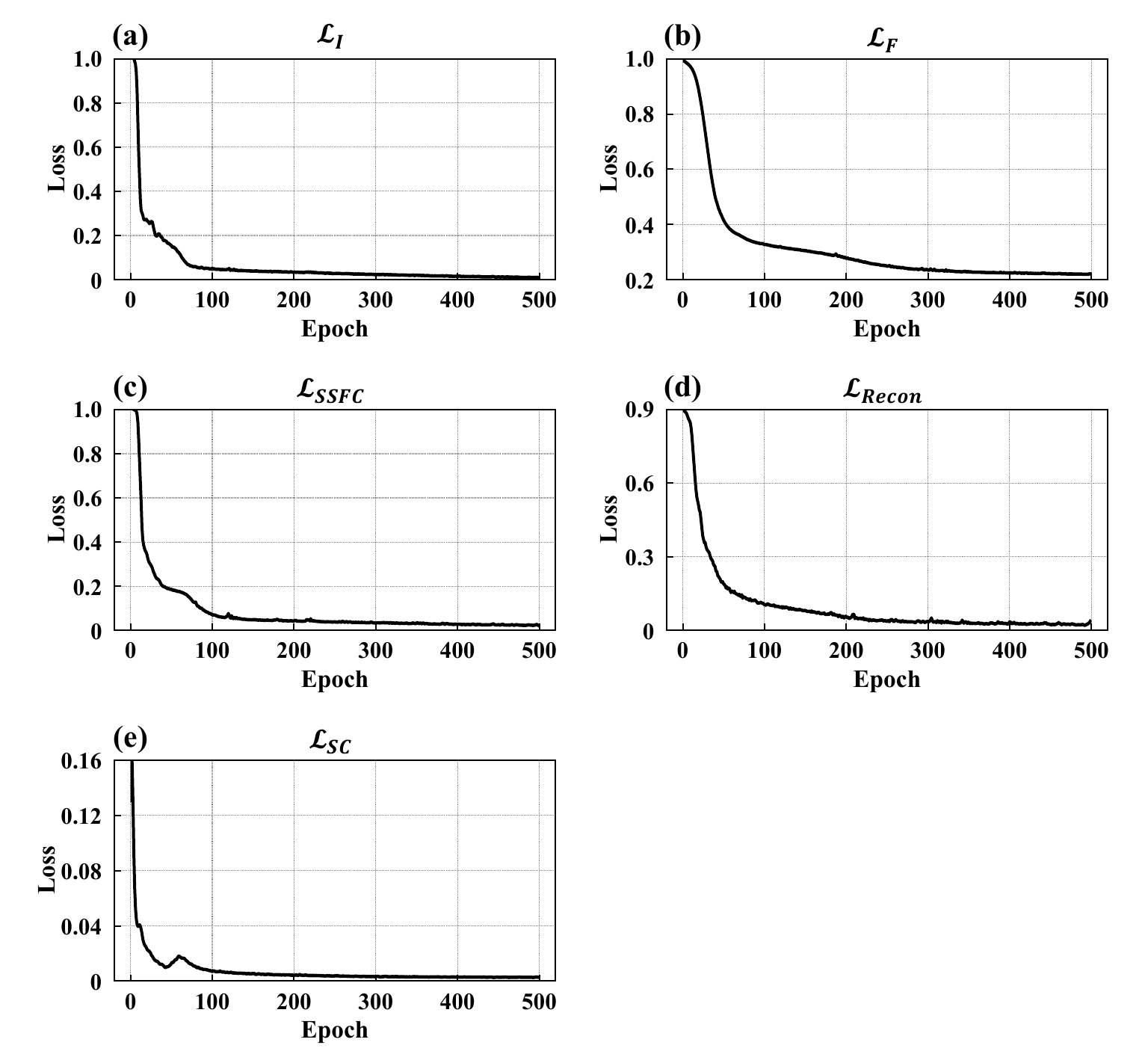}
\caption{Loss function curves of (a) $\mathcal{L}_{I}$, (b) $\mathcal{L}_{F}$, (c) $\mathcal{L}_{SSFC}$, (d) $\mathcal{L}_{Recon}$, and (e) $\mathcal{L}_{SC}$.}
\label{fig9}
\end{figure}
\begin{figure*}[!t]
\centering
\includegraphics[width=7.2in]{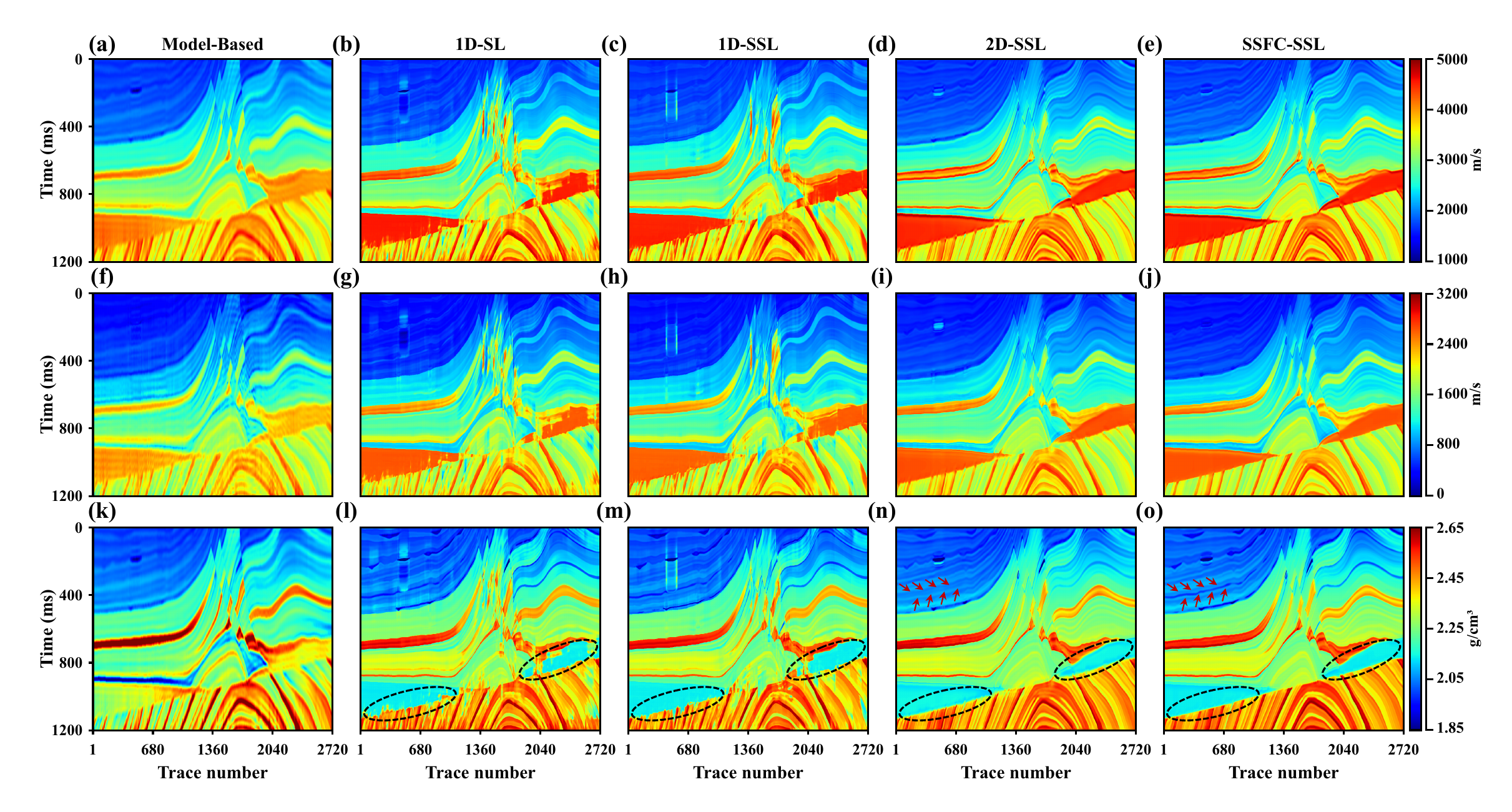}
\caption{Inversion results for (a)-(e) P-wave velocity, (f)-(j) S-wave velocity, and (k)-(o)
density using different methods. The first
column shows the model-based inversion results, the second column shows the 1D-SL inversion results, the third column shows the 1D-SSL inversion results, the fourth column shows the 2D-SSL inversion results, and the last column shows the SSFC-SSL inversion results.}
\label{fig10}
\end{figure*}
\begin{figure*}[!t]
\centering
\includegraphics[width=7.2in]{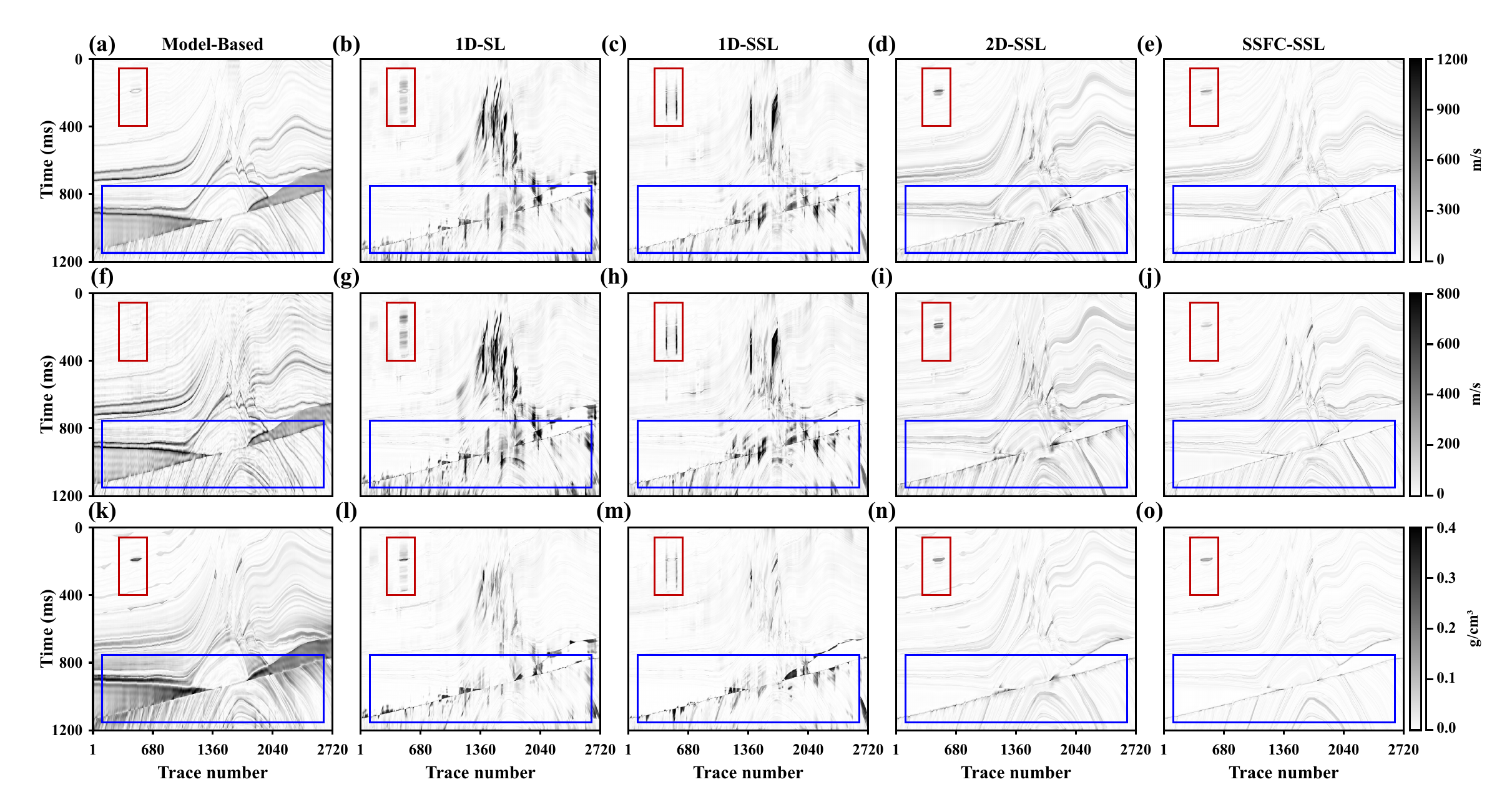}
\caption{Absolute difference in (a-e) P-wave velocity, (f-j) S-wave velocity, and (k-o)
density inversion results compared to the true data. The figure layout is consistent
with Fig. \ref{fig10}.}
\label{fig11}
\end{figure*}
\begin{figure*}[!t]
\centering
\includegraphics[width=7.2in]{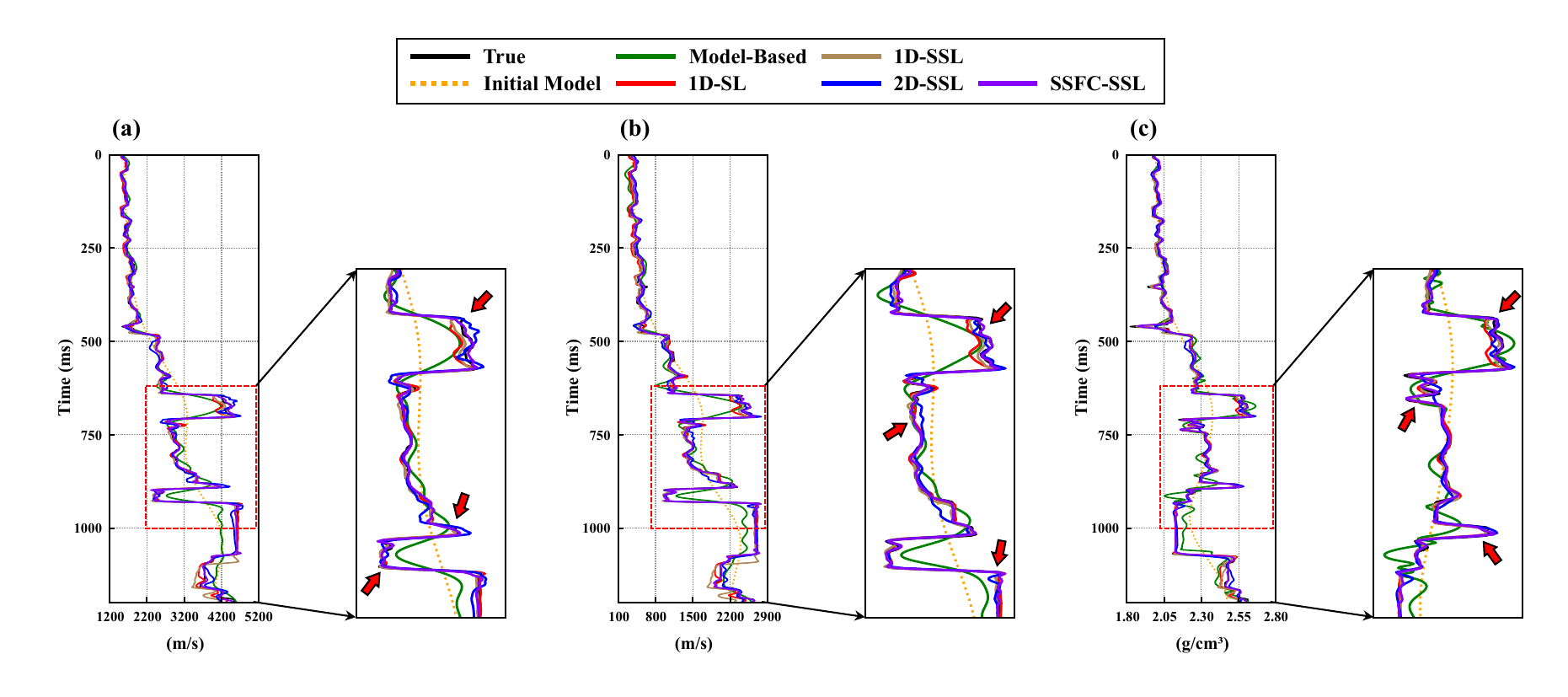}
\caption{Inversion results of (a) P-wave velocity, (b) S-wave velocity, and (c) density at trace 600 on the synthetic data. The black lines indicate true data, the orange dotted lines indicate the initial models, the green lines indicate the model-based inversion results, the red lines indicate the 1D-SL inversion results, the brown lines indicate the 1D-SSL inversion results, the blue lines indicate the 2D-SSL inversion results, and the purple lines indicate the SSFC-SSL inversion results.}
\label{fig12}
\end{figure*}

\begin{figure}[!t]
\centering
\includegraphics[width=3.5in]{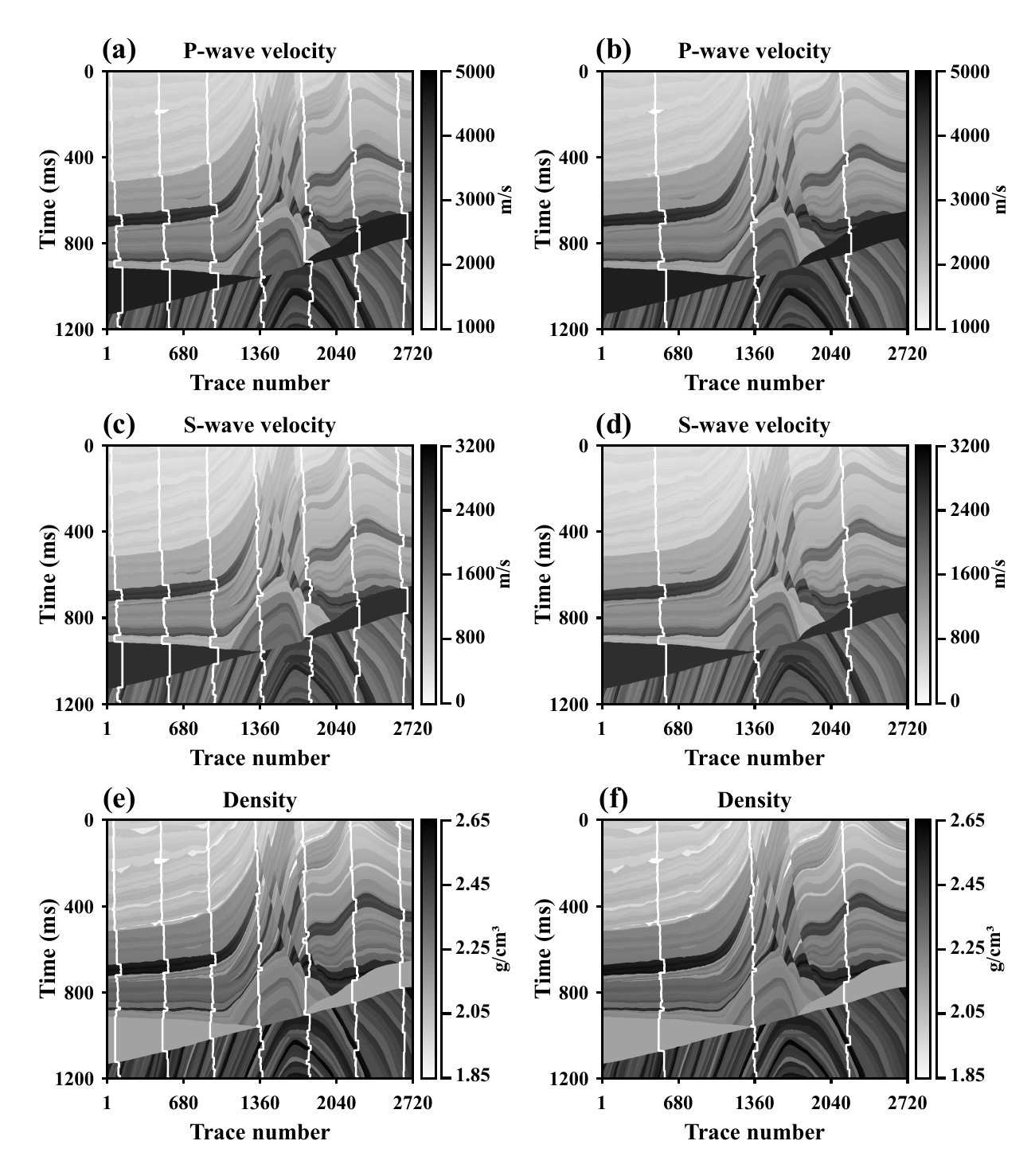}
\caption{(a) Seven training traces and (b) three training traces for P-wave velocity; (c) seven training traces and (d) three training traces for S-wave velocity; (e) seven training traces and (f) three training traces for density.}
\label{fig13}
\end{figure}

\begin{table}[!t]
\caption{Hyperparameters in SSFC-SSL.}
\renewcommand\arraystretch{1.2}
\centering
\label{table1}
\begin{tabular}{ll|c}
\toprule[1.5pt]
\multicolumn{2}{c|}{\textbf{The type of the hyperparameter}}                                 & \textbf{Value}    \\ \hline
\multicolumn{1}{l|}{\multirow{6}{*}{\textbf{Basic parameter}}}   & Epoch                     & 500               \\ \cline{2-3} 
\multicolumn{1}{l|}{}                                            & Batch size                & 100               \\ \cline{2-3} 
\multicolumn{1}{l|}{}                                            & Learning rate             & 0.002             \\ \cline{2-3} 
\multicolumn{1}{l|}{}                                            & Weight decay              & 0.0001            \\ \cline{2-3} 
\multicolumn{1}{l|}{}                                            & Dropout                   & 0.2               \\ \cline{2-3} 
\multicolumn{1}{l|}{}                                            & Initial kernel size       & 3                 \\ \hline
\multicolumn{1}{l|}{\multirow{3}{*}{\textbf{Inversion network}}} & Number of channels of TCN & {[}16,32,32,16{]} \\ \cline{2-3} 
\multicolumn{1}{l|}{}                                            & Dilation factor of TCN    & {[}1,2,4,8{]}     \\ \cline{2-3} 
\multicolumn{1}{l|}{}                                            & Number of channels of CNN & {[}32,16,3{]}     \\ \hline
\multicolumn{1}{l|}{\textbf{Forward network}}                    & Number of channels of CNN & {[}16,32,16,5{]}  \\ \hline
\multicolumn{1}{l|}{\textbf{SSFC network}}                       & Number of channels of CNN & {[}16,32,16,3{]}  \\ \bottomrule[1.5pt]
\end{tabular}
\end{table}

\begin{figure*}[!t]
\centering
\includegraphics[width=7.2in]{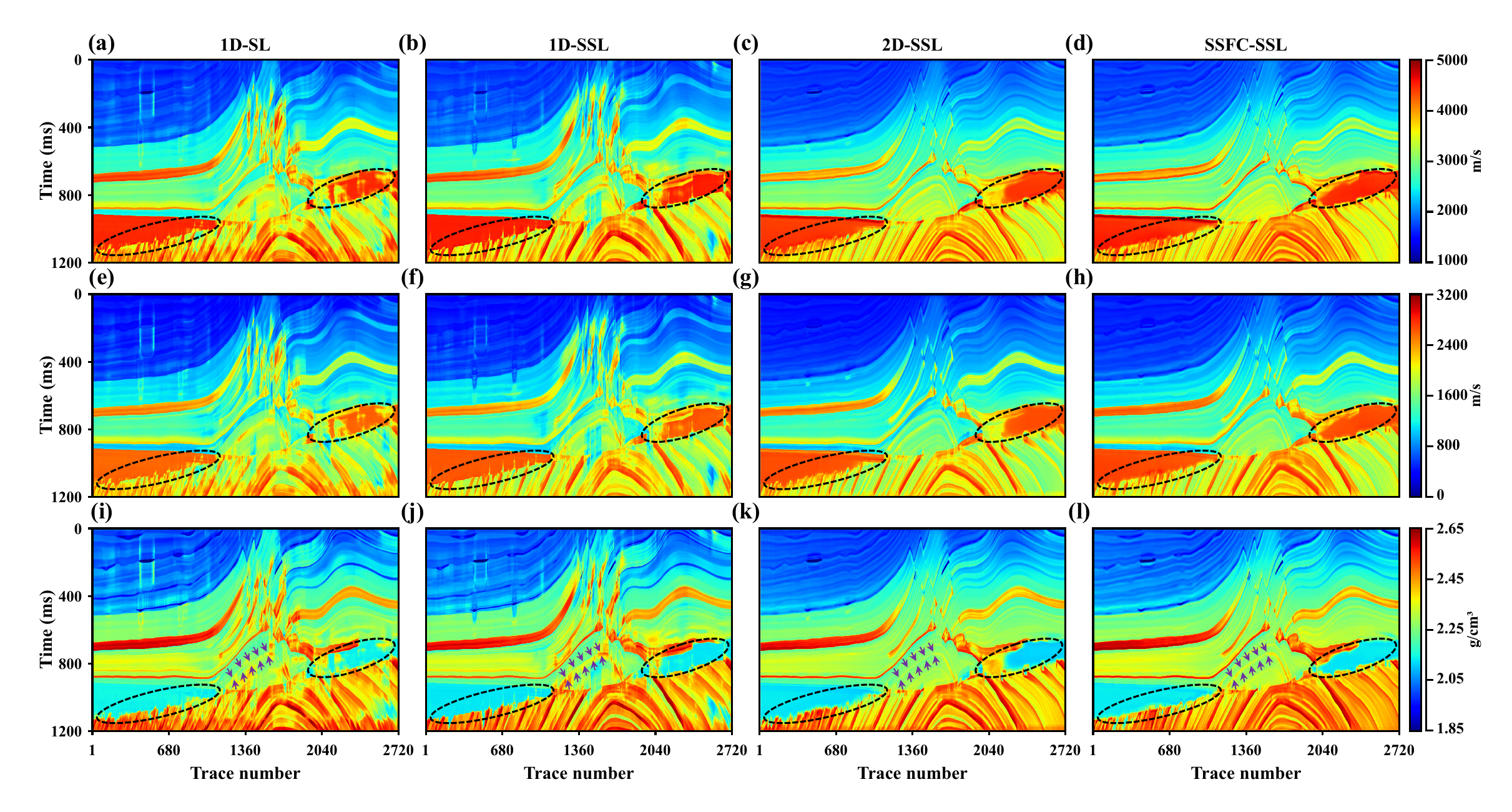}
\caption{Inversion results for (a)-(e) P-wave velocity, (f)-(j) S-wave velocity, and (k)-(o)
density using different methods with seven training traces. The first column shows the 1D-SL inversion results, the second column shows the 1D-SSL inversion results, the third column shows the 2D-SSL inversion results, and the last column shows the SSFC-SSL inversion results.}
\label{fig14}
\end{figure*}

\begin{figure*}[!t]
\centering
\includegraphics[width=7.2in]{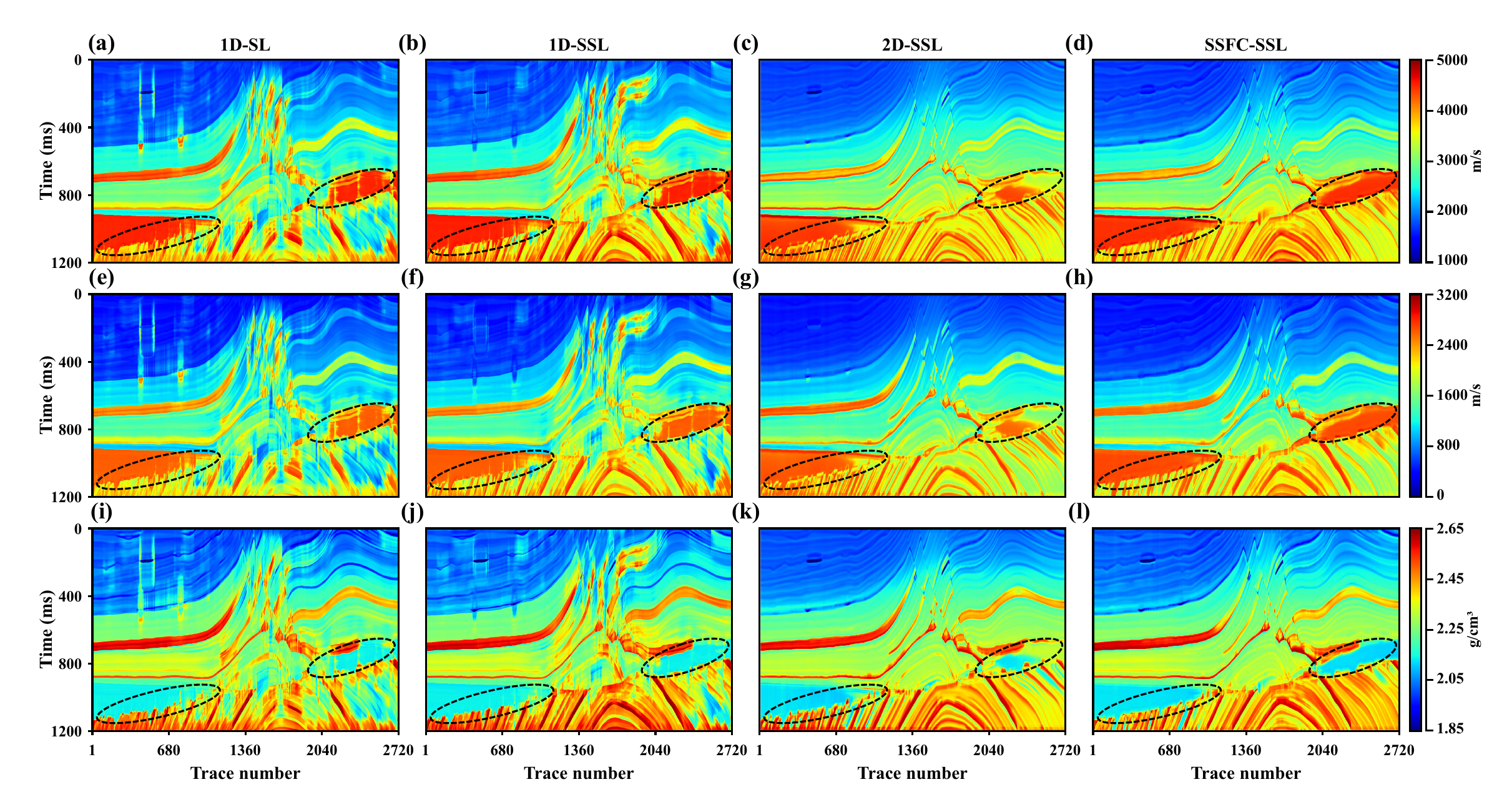}
\caption{Inversion results for (a)-(e) P-wave velocity, (f)-(j) S-wave velocity, and (k)-(o)
density using different methods with three training traces. The figure layout is consistent
with Fig. \ref{fig14}.}
\label{fig15}
\end{figure*}

\begin{figure*}[!t]
\centering
\includegraphics[width=6in]{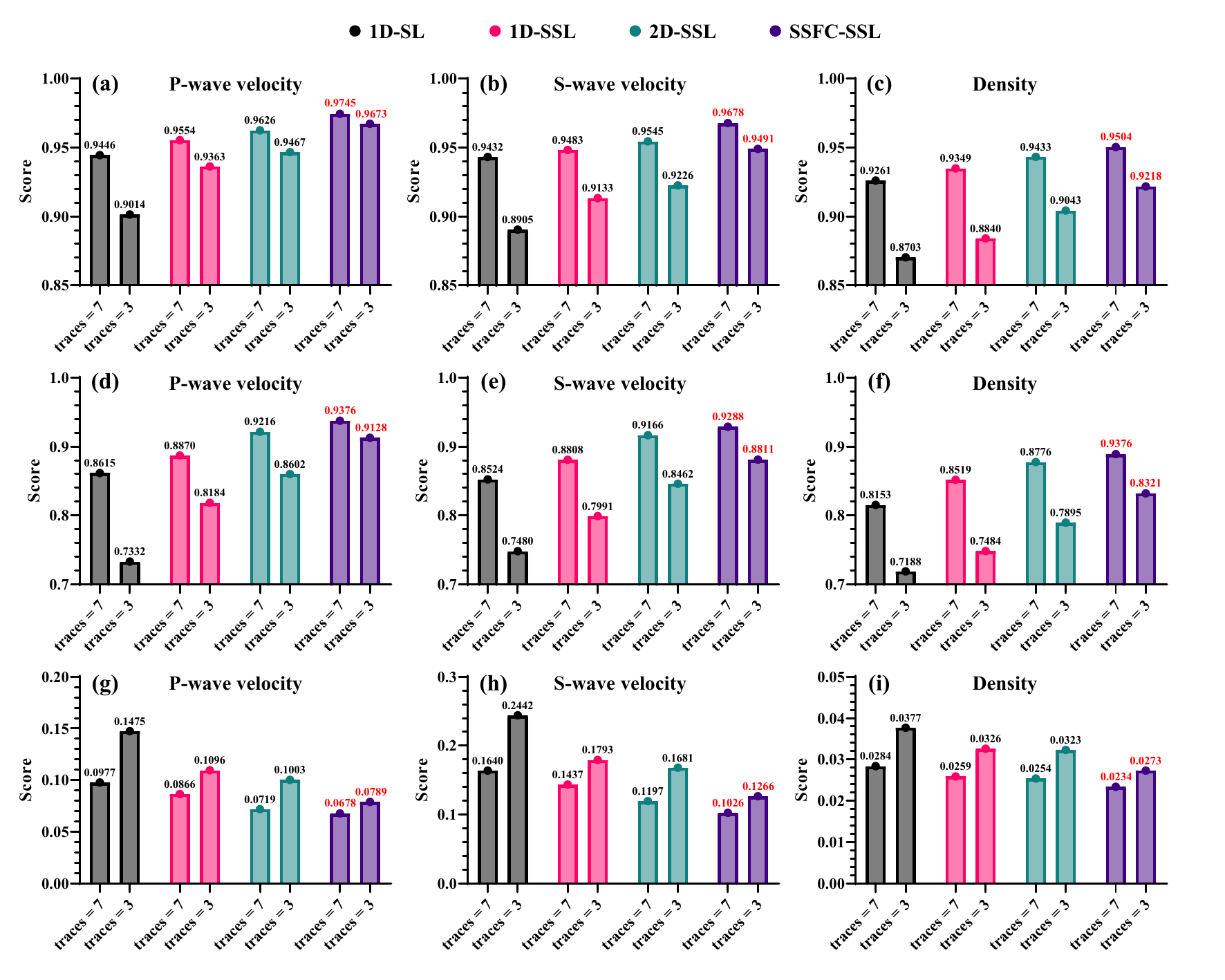}
\caption{Comparison of the (a)-(c) $\mathrm{PCC}$, (d)-(f) $\mathrm{R^{2}}$, and (g)-(i) $\mathrm{RMSE}$ of the different methods.}
\label{fig16}
\end{figure*}

\begin{figure*}[!t]
\centering
\includegraphics[width=7in]{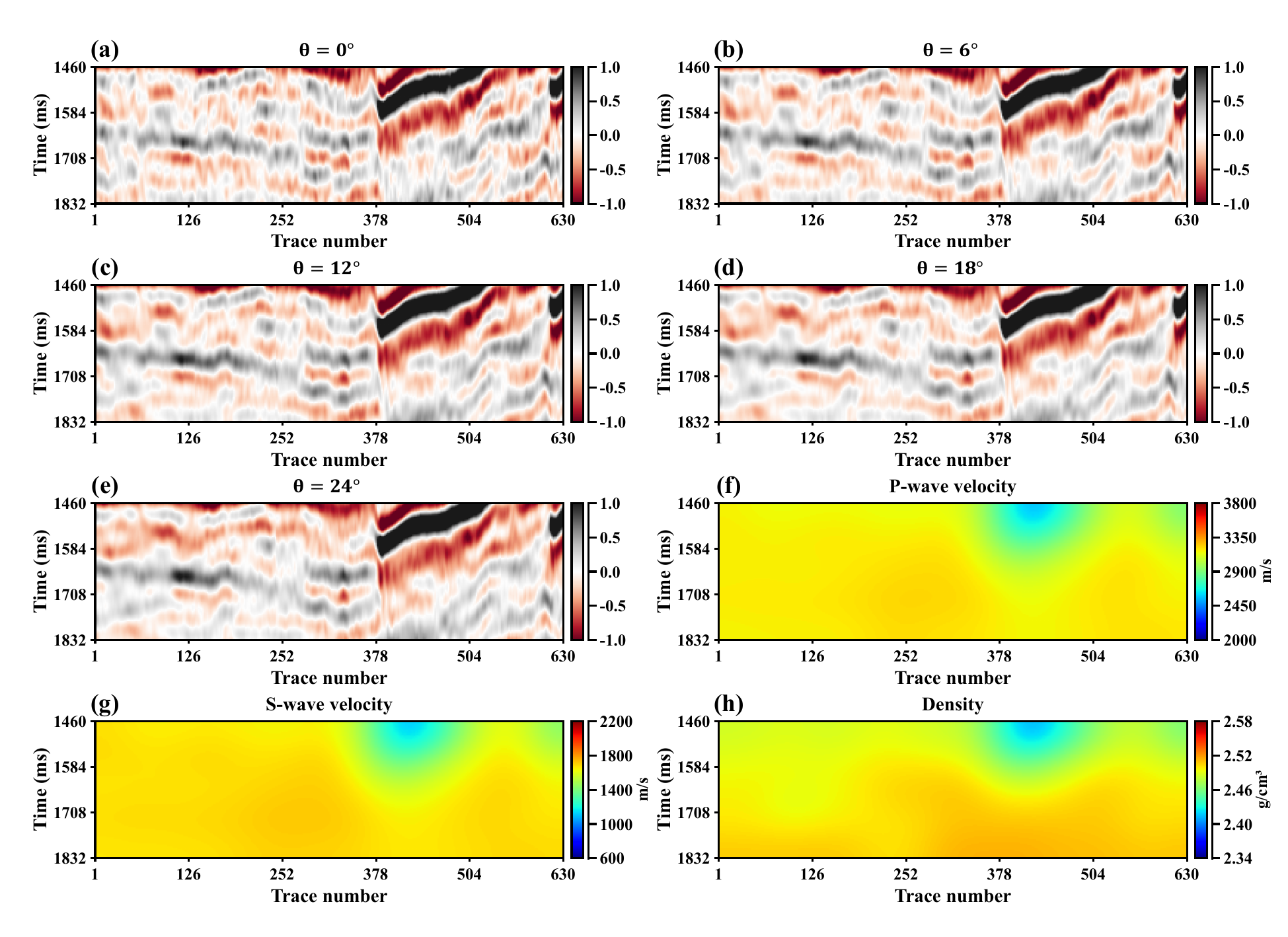}
\caption{Field seismic profiles of (a) 0$^{\circ}$, (b) 6$^{\circ}$, (c) 12$^{\circ}$, (d) 18$^{\circ}$, and (e) 24$^{\circ}$. Low-frequency profile of (f) P-wave velocity, (g) S-wave velocity, and (h) density.}
\label{fig17}
\end{figure*}

AVO is a crucial technique for characterizing subsurface reservoirs by analyzing the variation in seismic wave amplitudes with offset. Its theoretical foundation is the convolution model. In the absence of random noise, the convolution model can be expressed as
\begin{equation}\label{H1}
D = W*R_{PP}
\end{equation}
where $D$ represents seismic data, represented as the matrix $M*N$, where $M$ denotes the number of traces and 
$N$ represents the depth of a single trace sequence, $f$ represents the mapping matrix between seismic parameters and reflection coefficients, $W$ is the seismic wavelet, and $R_{pp}$ denotes the P-wave reflection coefficient, which is typically a function of velocity, density, and incident angle. Zoeppritz (1919) proposed equations to describe the reflection and transmission coefficients of PP and PS waves generated when a longitudinal wave in an isotropic medium meets a reflective interface. The equations are

\begin{equation}\label{H2}
\begin{split}
&
\begin{bmatrix} 
R_{PP} \\
R_{PS} \\
T_{PP} \\
T_{PP}
\end{bmatrix}\\
&=
\setlength{\arraycolsep}{0pt}
\begin{bmatrix} 
\sin\theta_{1} & \cos\varphi_{1} & -\sin\theta_{2} & \cos\varphi_{2} \\  
-\cos\theta_{1}  & \sin\varphi_{1} & -\cos\theta_{2} & -\sin\varphi_{2}\\
\sin2\theta_{1}  & \frac{V_{P_{1}}}{V_{S_{1}}} \cos2\varphi_{1} & \frac{\rho_{2}V_{S_{2}}^{2}V_{P_{1}}}{\rho_{1}V_{S_{1}}^{2}V_{P_{2}}} \sin2\theta_{2} & \frac{-\rho_{2}V_{S_{2}}V_{P_{1}}}{\rho_{1}V_{S_{1}}^{2} } \cos2\varphi_{2}\\ 
\cos2\varphi_{1}  & \frac{-V_{S_{1}}}{V_{P_{1}}} \sin2\varphi_{1} & \frac{-\rho_{2}V_{P_{2}}}{\rho_{1}V_{P_{1}}} \cos2\varphi_{2} & \frac{-\rho_{2}V_{S_{2}}}{\rho_{1}V_{P_{1}}} \sin2\varphi_{2}
\end{bmatrix}^{-1} \\
&\times
\begin{bmatrix} 
-\sin\theta_{1} \\
-\cos\theta_{1} \\
\sin2\theta_{1} \\
-\cos2\varphi_{1}
\end{bmatrix}
\end{split}
\end{equation}
where $\theta_{1}$, $\theta_{2}$, $\varphi_{1}$, and $\varphi_{2}$ represent the incident P-wave angle, the transmitted P-wave angle, the reflected S-wave angle, and the transmitted S-wave angle, respectively. $V_{P_{1}}$, $V_{S_{1}}$, and $\rho_{1}$ represent the P-wave velocity, S-velocity, and density of the medium in the upper layer of the reflective interface, while $V_{P_{2}}$, $V_{S_{2}}$, and $\rho_{2}$ represent the P-wave velocity, S-wave velocity, and density of the medium in the lower layer. $R_{PP}$, $R_{PS}$, $T_{PP}$, and $T_{PS}$ correspond to the coefficients of the reflected P-wave, reflected S-wave, transmitted P-wave, and transmitted S-wave, respectively. According to Snell's law, the relationship between the incident angle and velocity can be expressed as 
\begin{equation}\label{H3}
\frac{\sin\theta_{1}}{V_{P_{1}}} =\frac{\sin\theta_{2}}{V_{P_{2}}} 
=\frac{\sin\varphi_{1}}{V_{S_{1}}}
=\frac{\sin\varphi_{2}}{V_{S_{2}}}
\end{equation}
By combining equations (\ref{H3}) and (\ref{H2}), the reflection coefficient can be simplified as
\begin{equation}\label{H4}
R_{PP} = f(m)
\end{equation}
where $m$ is a vector set of elastic parameters: P-wave velocity, S-wave velocity, and density, represented as $m=[V_P, V_S, \rho]$, $f$ represents the mapping matrix between seismic parameters and reflection coefficients.

By substituting (\ref{H4}) into (\ref{H1}), the general form of the forward equation in AVO theory is obtained, expressed as
\begin{equation}\label{H5}
D = W*f(m) = G(m)
\end{equation}
where $G$ represents the forward operator.

AVO inversion is the reverse process of AVO forward modeling. When the mapping matrix $G$ is invertible and the influence of the seismic wavelet is disregarded, the inversion result can be expressed as
\begin{equation}\label{H6}
m = G^{-1}(D)
\end{equation}
where $G^{-1}$ represents the inverse operator of $G$. In AVO inversion, the mapping operator 
$G^{-1}$ is typically non-invertible.
\subsection{Semi-supervised seismic inversion structure}
When deep learning is applied to AVO inversion, the mapping operator $G^{-1}$ in equation (\ref{H6}) is replaced by a network, mathematically expressed as 
\begin{equation}\label{H7}
m=\mathcal{I}_{w}(D)
\end{equation}
where $\mathcal{I}_{w}$ denotes the inversion network, and $w$ denotes the trainable network parameters. The network can perform supervised learning using well-log data, with the objective function expressed as 
\begin{equation}\label{H8}
J(\mathcal{I}_{w}) = 
\min_{\mathcal{I}_{w}}||{m_{l}}-\mathcal{I}_{w}(D_{l})||_2^2
\end{equation}
where where  $\left \| * \right \|_{2}^{2}$ denotes the L2 norm, $m_{l}$ represents the elastic parameters from well-log and $D_{l}$ denotes the labeled seismic data. 

However, the proportion of well-log data is extremely small compared to the entire seismic dataset. This leads to a significant shortage of training data, which significantly affects inversion accuracy and limits the practical application of this method. To address this issue, we constructed a forward network, expressed as 
\begin{equation}\label{H9}
D=\mathcal{F}_{w}(m)
\end{equation}
where $\mathcal{F}_{w}$ represents the forward network. 
For supervised training with well-log data, its objective function is
\begin{equation}\label{H10}
J(\mathcal{F}_{w}) = 
\min_{\mathcal{F}_{w}}||{D_{l}}-\mathcal{F}_{w}(m_{l})||_2^2
\end{equation}

Moreover, to improve the generalization ability of the inversion and forward networks, we process a large amount of unlabeled seismic data (i.e., seismic data not from well locations) through the inversion and forward networks. The objective function is expressed as 
\begin{equation}\label{H11}
J(\mathcal{I}_{w},\mathcal{F}_{w}) = 
\min_{\mathcal{I}_{w},\mathcal{F}_{w}}||{D_{u}}-\mathcal{F}_{w}(\mathcal{I}_{w}(D_{u}))||_2^2
\end{equation}
where $D_{u}$ represents the unlabeled seismic data. Finally, by combining equations (\ref{H8}), (\ref{H10}), and (\ref{H11}), the objective function is obtained: 
\begin{equation}\label{H12}
\begin{split}
J(\mathcal{I}_{w},\mathcal{F}_{w}) 
&= 
\min_{\mathcal{I}_{w}}||{m_{l}}-\mathcal{I}_{w}(D_{l})||_2^2 
\\
&+ \min_{\mathcal{F}_{w}}||{D_{l}}-\mathcal{F}_{w}(m_{l})||_2^2
\\
&+ \min_{\mathcal{I}_{w},\mathcal{F}_{w}}||{D_{u}}-\mathcal{F}_{w}(\mathcal{I}_{w}(D_{u}))||_2^2
\end{split}
\end{equation}
\subsection{Strong spatial feature constraints
}
% In practical applications, well-log data is typically used to establish the relationship between single-trace prediction parameters and well-log parameters, while the correlation between adjacent traces is often ignored. 
To establish the mapping of strong spatial features between the predicted parameters at well locations and those at non-well locations, we propose a strong spatial feature constraint method. The matrix forms of $m_l$ and $\mathcal{I}_{w}(D_{l})$ in (\ref{H11}) are given by
\begin{equation}\label{H13}
\begin{split}
&m_{l}=
\begin{bmatrix}  0&\dots&m_{1+\frac{M-1}{2},1}&\dots&0\\ \vdots & \ddots  & \vdots & \ddots & \vdots\\  0&\dots&m_{1+\frac{M-1}{2},N}&\dots&0\end{bmatrix}_{N \times M}
\\
&\mathcal{I}_{w}(D_{l})=
\\
&\begin{bmatrix}  \widehat{m} _{1,1}&\dots&\widehat{m}_{1+\frac{M-1}{2},1}&\dots&\widehat{m}_{M,1}\\ \vdots & \ddots  & \vdots & \ddots & \vdots\\  \widehat{m}_{1,N}&\dots&\widehat{m}_{1+\frac{M-1}{2},N}&\dots&\widehat{m}_{M,N}\end{bmatrix}_{N \times M}
\end{split}
\end{equation}
where $\widehat{m}$ represents the predicted parameters. It can be seen from equation (\ref{H13}) that the well-log label only constraints the $1+\frac{M-1}{2}$th trace, which corresponds to the middle trace. To enable the well-log label to constrain adjacent traces, we define a label ablation factor, expressed as
\begin{equation}\label{H14}
R=\begin{bmatrix}  1&\dots&0&\dots&1\\ \vdots & \ddots  & \vdots & \ddots & \vdots\\  1&\dots&0&\dots&1\end{bmatrix}_{N \times M}
\end{equation}
We can set the predicted parameters of the middle trace to zero by using $\widehat{m}*R$. Furthermore, a strong spatial feature constraint network $\mathcal{S}_{w}$ was constructed to recover the missing data of the middle trace using adjacent traces. To ensure the stable convergence of the network, the self-consistency \cite{pourpanah2022review}, \cite{wang2022self}, \cite{peng2024zero} is introduced. The objective function is expressed as
\begin{equation}\label{H15}
\begin{split}
J(\mathcal{S}_{w}) 
&= 
\min_{\mathcal{S}_{w}}||m_{l}-(\mathcal{S}_{w}({\widehat{m}_{l}*R}))*(1-R)||_2^2 
\\
&+ \min_{\mathcal{S}_{w}}||{\widehat{m}_{l}*R}-(\mathcal{S}_{w}({\widehat{m}_{l}*R}))*R||_2^2
\end{split}
\end{equation}
By combining equations (\ref{H12}) and (\ref{H15}), the inversion objective function is obtained:
\begin{equation}\label{H16}
\begin{split}
J(\mathcal{I}_{w},\mathcal{F}_{w},\mathcal{S}_{w})
&= 
\min_{\mathcal{I}_{w}}||{m_{l}}-\mathcal{I}_{w}(D_{l})||_2^2 
\\
&+ \min_{\mathcal{F}_{w}}||{D_{l}}-\mathcal{F}_{w}(m_{l})||_2^2
\\
&+ \min_{\mathcal{I}_{w},\mathcal{F}_{w}}||{D_{u}}-\mathcal{F}_{w}(\mathcal{I}_{w}(D_{u}))||_2^2
\\
&+
\min_{\mathcal{S}_{w}}||m_{l}-(\mathcal{S}_{w}({\widehat{m}_{l}*R}))*(1-R)||_2^2
\\
&+ \min_{\mathcal{S}_{w}}||{\widehat{m}_{l}*R}-(\mathcal{S}_{w}({\widehat{m}_{l}*R}))*R||_2^2
\end{split}
\end{equation}
In addition, SSFC-SSL uses the low-frequency information of the elastic parameters as conditional input to improve the inversion accuracy. Equation (\ref{H16}) is written as 
\begin{equation}\label{H17}
\begin{split}
J(\mathcal{I}_{w},\mathcal{F}_{w},\mathcal{S}_{w})
&= 
\min_{\mathcal{I}_{w}}\underbrace{||{m_{l}}-\mathcal{I}_{w}(D_{l},m_{low})||_2^2}_{\mathcal{L}_{I}} 
\\
&+ \min_{\mathcal{F}_{w}}\underbrace{||{D_{l}}-\mathcal{F}_{w}(m_{l})||_2^2}_{\mathcal{L}_{F}}
\\
&+ \min_{\mathcal{I}_{w},\mathcal{F}_{w}}\underbrace{||{D_{u}}-\mathcal{F}_{w}(\mathcal{I}_{w}(D_{u},m_{low}))||_2^2}_{\mathcal{L}_{Recon}}
\\
&+
\min_{\mathcal{S}_{w}}\underbrace{||m_{l}-(\mathcal{S}_{w}({\widehat{m}_{l}*R}))*(1-R)||_2^2}_{\mathcal{L}_{SSFC}}
\\
&+ \min_{\mathcal{S}_{w}}\underbrace{||{\widehat{m}_{l}*R}-(\mathcal{S}_{w}({\widehat{m}_{l}*R}))*R||_2^2}_{\mathcal{L}_{SC}}
\end{split}
\end{equation}
Therefore, the total loss of the SSFC-SSL network can be expressed as
\begin{equation}\label{H18}
Loss = \mathcal{L}_{I}+\mathcal{L}_{F}+\mathcal{L}_{SSFC}+\mathcal{L}_{Recon}+\mathcal{L}_{sc}
\end{equation}
where $\mathcal{L}_{I}$ represents the training loss of the inversion network, $\mathcal{L}_{F}$ represents the training loss of the forward network, $\mathcal{L}_{SSFC}$ represents the training loss of the SSFC network, $\mathcal{L}_{Recon}$ represents the loss from the reconstruction of unlabeled seismic data, and $\mathcal{L}_{SC}$ represents the self-consistency loss of the SSFC network.

% \begin{multline}\label{h12} %multline
% J(\mathcal{F}_{w},\mathcal{H}_{w}) = 
% \min_{\mathcal{F}_{w}} 
% \underbrace{||{I_{l}}-\mathcal{F}_{w}(D_{l})||_2^2}_{Loss_{I}} +
% \min_{\mathcal{H}_{w}} 
% \underbrace{||{D_{l}}-\mathcal{H}_{w}(I_{l})||_2^2}_{Loss_{C}} +
% \\
% \min_{\mathcal{F}_{w},\mathcal{H}_{w}} 
% \underbrace{||D_{U}-\mathcal{H}_{w}(\mathcal{F}_{w}(D_{U}))||_2^2}_{Loss_{T}}  + 
% \min_{\mathcal{F}_{w},\mathcal{H}_{w}} 
% \underbrace{||F_{U}-G^{\prime}\mathcal{H}_{w}(\mathcal{F}_{w}(D_{U}))||_2^2}_{Loss_{F}}  + 
% \\
% \min_{\mathcal{F}_{w},\mathcal{H}_{w}} 
% \underbrace{||P_{U}-K\mathcal{H}_{w}(\mathcal{F}_{w}(D_{U}))||_2^2}_{Loss_{P}},
% \end{multline}

\subsection{Workflow of SSFC-SSL}
The workflow of the SSFC-SSL method for AVO inversion consists of two parts: training and prediction. During the training phase, the SSFC-SSL method requires the simultaneous training of the forward-modeling network, the inversion network, and the SSFC network, thereby achieving closed-loop optimization among all three. 

Before starting the training, we initialize the weights and biases of the neural networks. The data include labeled, unlabeled seismic data, and well-log elastic parameters (P-wave velocity, S-wave velocity, and density). Additionally, a low-frequency model is integrated into the network, alongside the seismic data, as prior information. During training, labeled seismic data is used to optimize the forward and inversion networks. 
During the training process, both labeled and unlabeled seismic data are utilized to optimize the inversion and forward networks. The output of the forward network is fed back to the inversion network in the form of pseudo-labels, enabling the effective utilization of unlabeled data. Meanwhile, well-log elastic parameters contribute to the optimization of the inversion network, forward network, and SSFC network. Eventually, a mapping between multi-trace seismic data and multi-trace elastic parameters is achieved. 
During the prediction stage, the SSFC network and the forward network are no longer involved, and the inversion task is completed solely by leveraging the efficient inference capability of the inversion network.
\subsection{Network structure}
The SSFC-SSL consists of three sub-networks: the inversion network, the forward network, and the strong spatial feature constraints network. To accomplish the inversion task, we adopted the CNN as the basic architecture. Given the need to extract global features from the shallow structure in the inversion network, which requires a larger receptive field, we introduced temporal convolutional networks (TCN) and a dilated convolution design. Dilated convolution allows for precise control of the receptive field by introducing a dilation factor, enabling the network to achieve a larger receptive field with fewer layers, thereby capturing historical information over a longer time scale. Fig. \ref{fig3} shows an example of the 5×5 dilated convolution kernel structure, with an initial receptive field of 5. By introducing dilation factors of 2 and 3, the receptive fields for a single-layer network are expanded to 11 and 17, respectively. The dilated convolution significantly enhances the ability of the network to extract shallow global features and reduces the parameter count. 

Fig. \ref{fig4} illustrates the inversion network, consisting of four TCN blocks and three CNN blocks, $K$ is the number of seismic profiles corresponding to different incidence angles. The structure of the TCN, shown within the dashed box in Fig. \ref{fig4}, consists of two branches. The first branch transforms the input, while the second applies a 1$\times$1 convolution operation. Then, the outputs of the two branches are added together to generate the final result. The residual connection enables information to flow across layers, effectively preventing the loss of information due to excessive depth. Compared to Recurrent Neural Networks (RNNs), the TCN addresses the issues of vanishing and exploding gradients. Figs. \ref{fig5} and \ref{fig6} illustrate the structures of the forward network and the SSFC network, both of which consist of four convolutional layers, each followed by a Batch Normalization (BN) layer and a ReLU activation layer. The two network structures are designed to maintain consistency.

\begin{table*}[!t]
\caption{Quantitative analysis of the inversion results for the synthetic data.}
\renewcommand\arraystretch{1.2}
\centering
\label{table2}
\begin{tabular}{c|ccc|ccc|ccc}
\toprule[1.5pt]
\multirow{2}{*}{\textbf{Method}} & \multicolumn{3}{c|}{\textbf{P-wave Velocity}}                                                      & \multicolumn{3}{c|}{\textbf{S-wave Velocity}}                                                      & \multicolumn{3}{c|}{\textbf{Density}}                                                              \\ \cline{2-10} 
                                 & \multicolumn{1}{c|}{PCC}             & \multicolumn{1}{c|}{R²} & RMSE            & \multicolumn{1}{c|}{PCC}             & \multicolumn{1}{c|}{R²} & RMSE            & \multicolumn{1}{c|}{PCC}             & \multicolumn{1}{c|}{R²} & RMSE            \\ \hline
Model-Based                      & \multicolumn{1}{c|}{0.9731}          & \multicolumn{1}{c|}{0.9447}               & 0.0757          & \multicolumn{1}{c|}{0.9715}          & \multicolumn{1}{c|}{0.9402}               & 0.1291          & \multicolumn{1}{c|}{0.8478}          & \multicolumn{1}{c|}{0.6754}               & 0.0437          \\ \hline
1D-SL                            & \multicolumn{1}{c|}{0.9692}          & \multicolumn{1}{c|}{0.9172}               & 0.0777          & \multicolumn{1}{c|}{0.9680}          & \multicolumn{1}{c|}{0.9143}               & 0.1288          & \multicolumn{1}{c|}{0.9491}          & \multicolumn{1}{c|}{0.8831}               & 0.0231          \\ \hline
1D-SSL                           & \multicolumn{1}{c|}{0.9786}          & \multicolumn{1}{c|}{0.9467}               & 0.0654          & \multicolumn{1}{c|}{0.9775}          & \multicolumn{1}{c|}{0.9442}               & 0.1093          & \multicolumn{1}{c|}{0.9597}          & \multicolumn{1}{c|}{0.9049}               & 0.0215          \\ \hline
2D-SSL                           & \multicolumn{1}{c|}{0.9846}          & \multicolumn{1}{c|}{0.9669}               & 0.0553          & \multicolumn{1}{c|}{0.9863}          & \multicolumn{1}{c|}{0.9637}               & 0.0832          & \multicolumn{1}{c|}{0.9751}          & \multicolumn{1}{c|}{0.9403}               & 0.0166          \\ \hline
SSFC-SSL                         & \multicolumn{1}{c|}{\textbf{0.9911}} & \multicolumn{1}{c|}{\textbf{0.9814}}      & \textbf{0.0420} & \multicolumn{1}{c|}{\textbf{0.9892}} & \multicolumn{1}{c|}{\textbf{0.9779}}      & \textbf{0.0624} & \multicolumn{1}{c|}{\textbf{0.9868}} & \multicolumn{1}{c|}{\textbf{0.9617}}      & \textbf{0.0123} \\ \bottomrule[1.5pt]
\end{tabular}
\end{table*}
\begin{figure}[!t]
\centering
\includegraphics[width=3.5in]{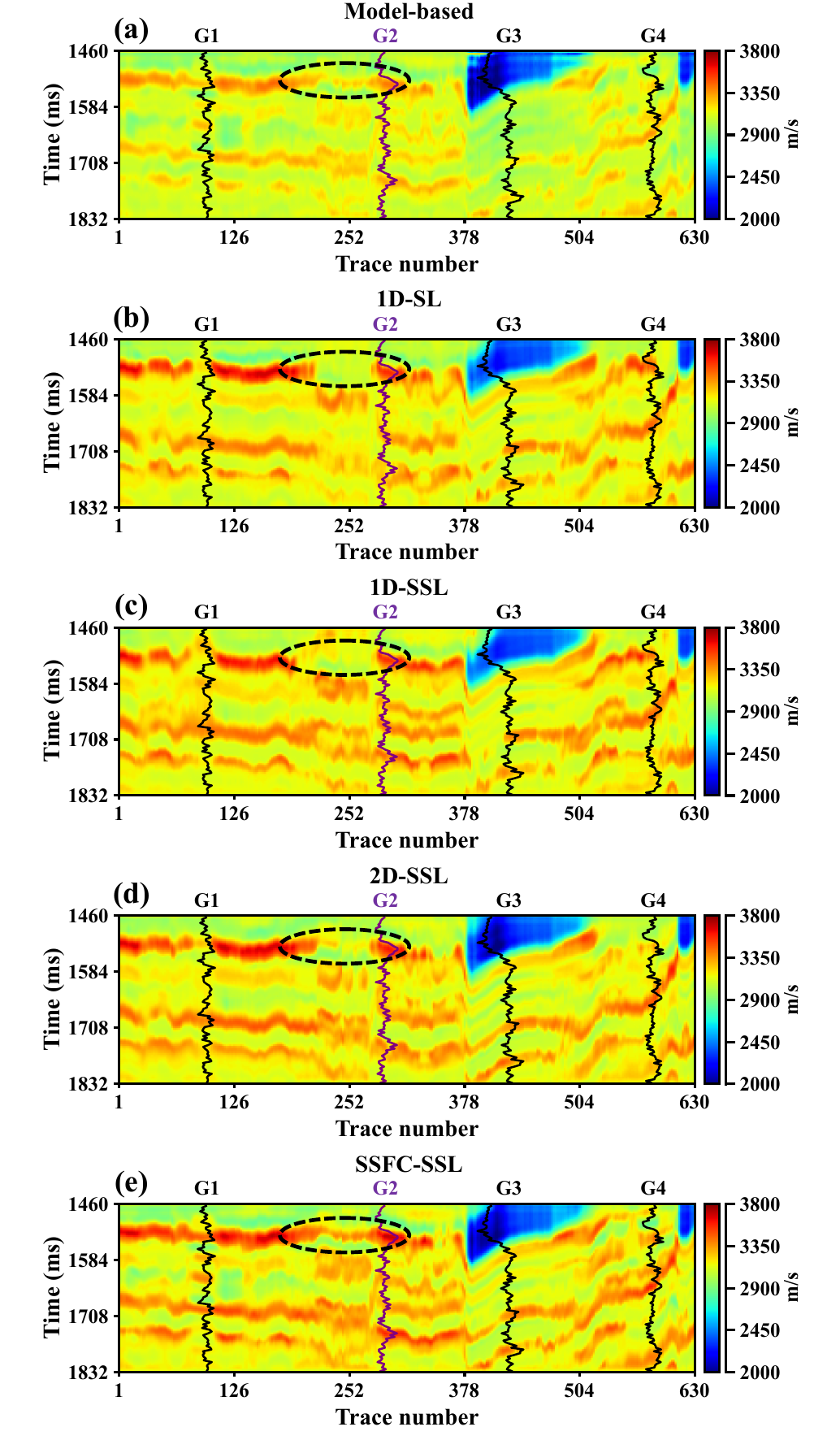}
\caption{Inversion results of P-wave velocity using (a) model-based, (b) 1D-SL, (c) 1D-SSL, (d) 2D-SSL, and (e) SSFC-SSL methods.}
\label{fig18}
\end{figure}
\begin{figure}[!t]
\centering
\includegraphics[width=3.5in]{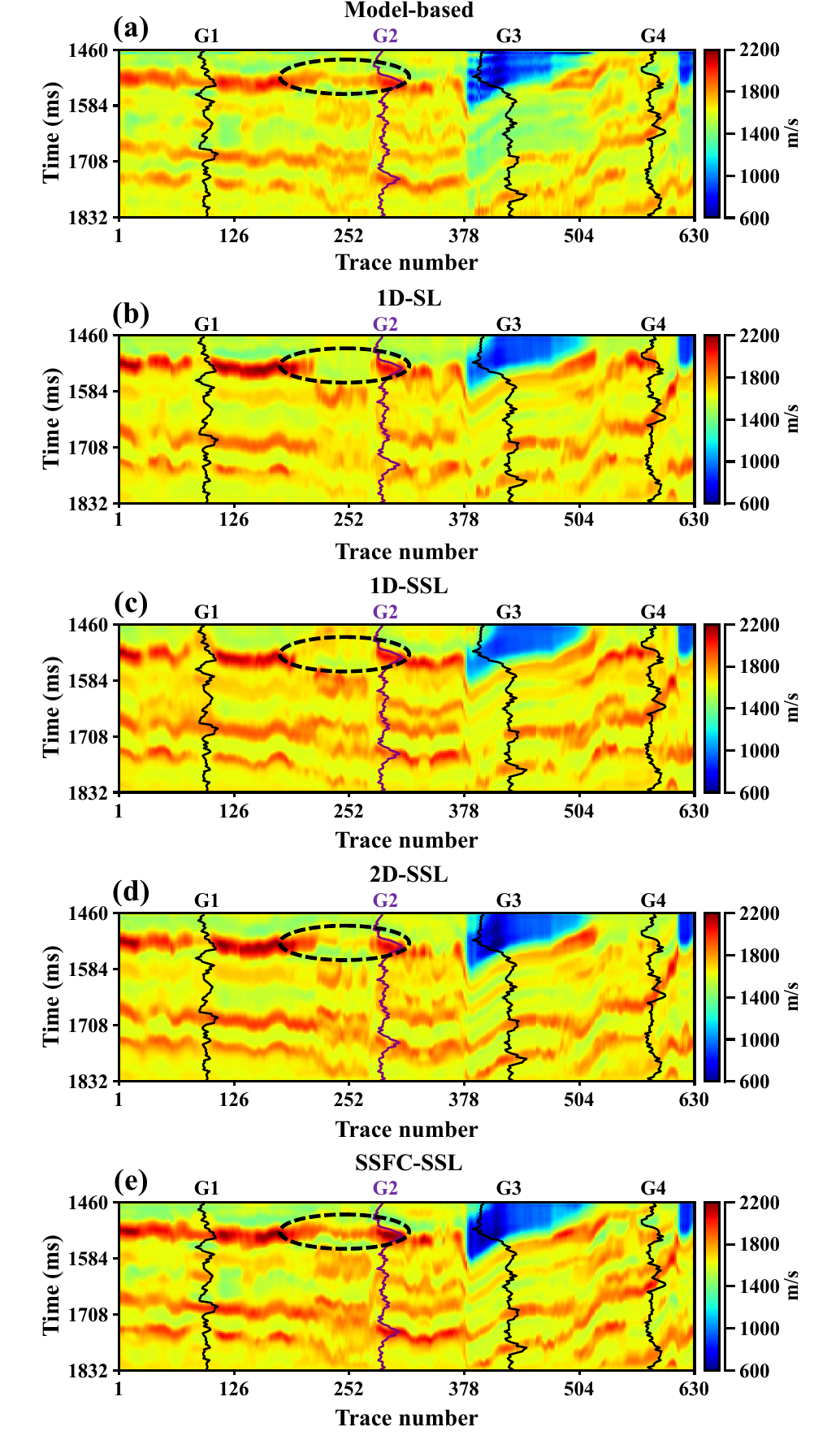}
\caption{Inversion results of S-wave velocity using (a) model-based, (b) 1D-SL, (c) 1D-SSL, (d) 2D-SSL, and (e) SSFC-SSL methods.}
\label{fig19}
\end{figure}
\begin{figure}[!t]
\centering
\includegraphics[width=3.5in]{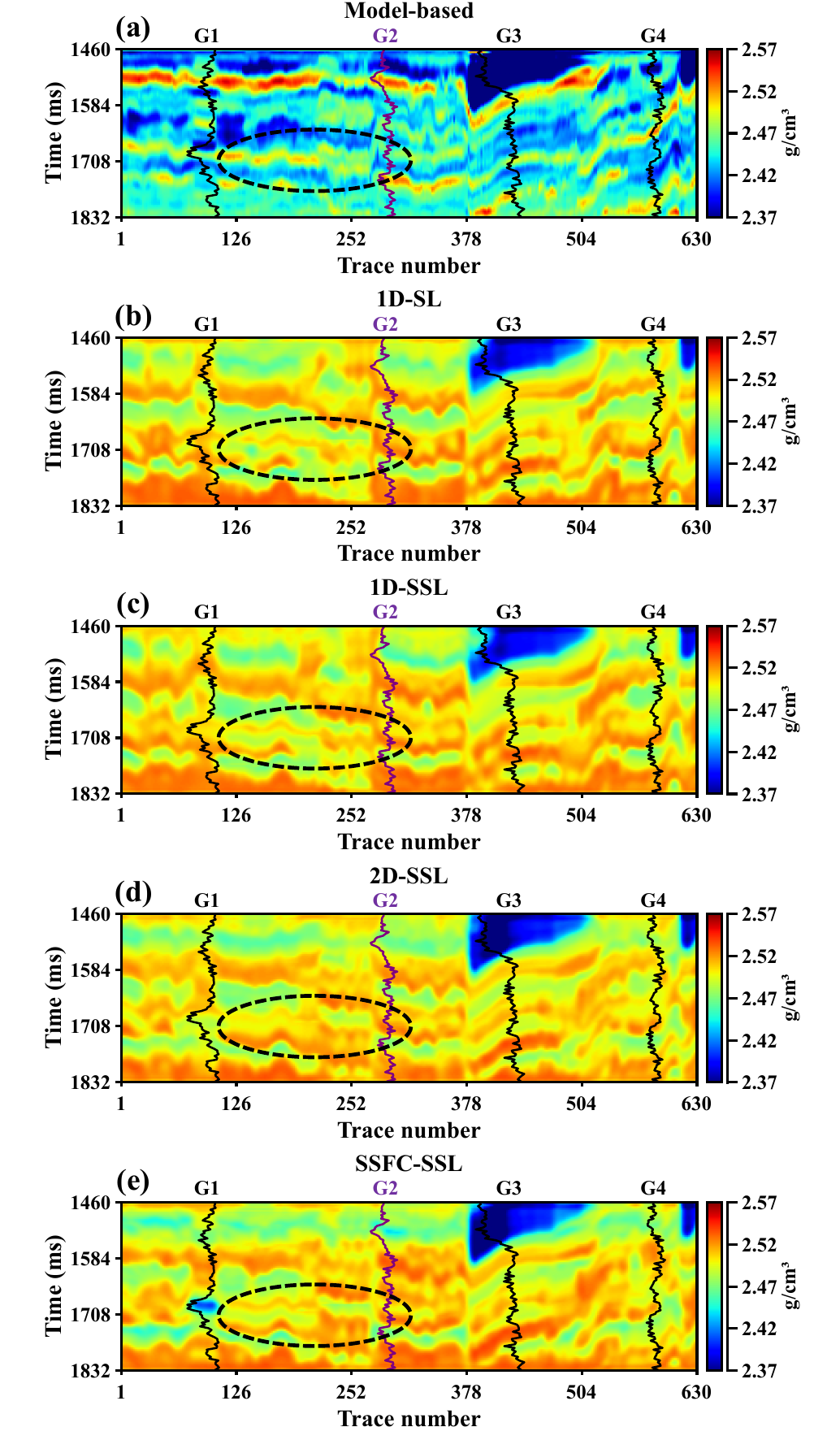}
\caption{Inversion results of density using (a) model-based, (b) 1D-SL, (c) 1D-SSL, (d) 2D-SSL, and (e) SSFC-SSL methods.}
\label{fig20}
\end{figure}

\begin{figure*}[!t]
\centering
\includegraphics[width=6.5in]{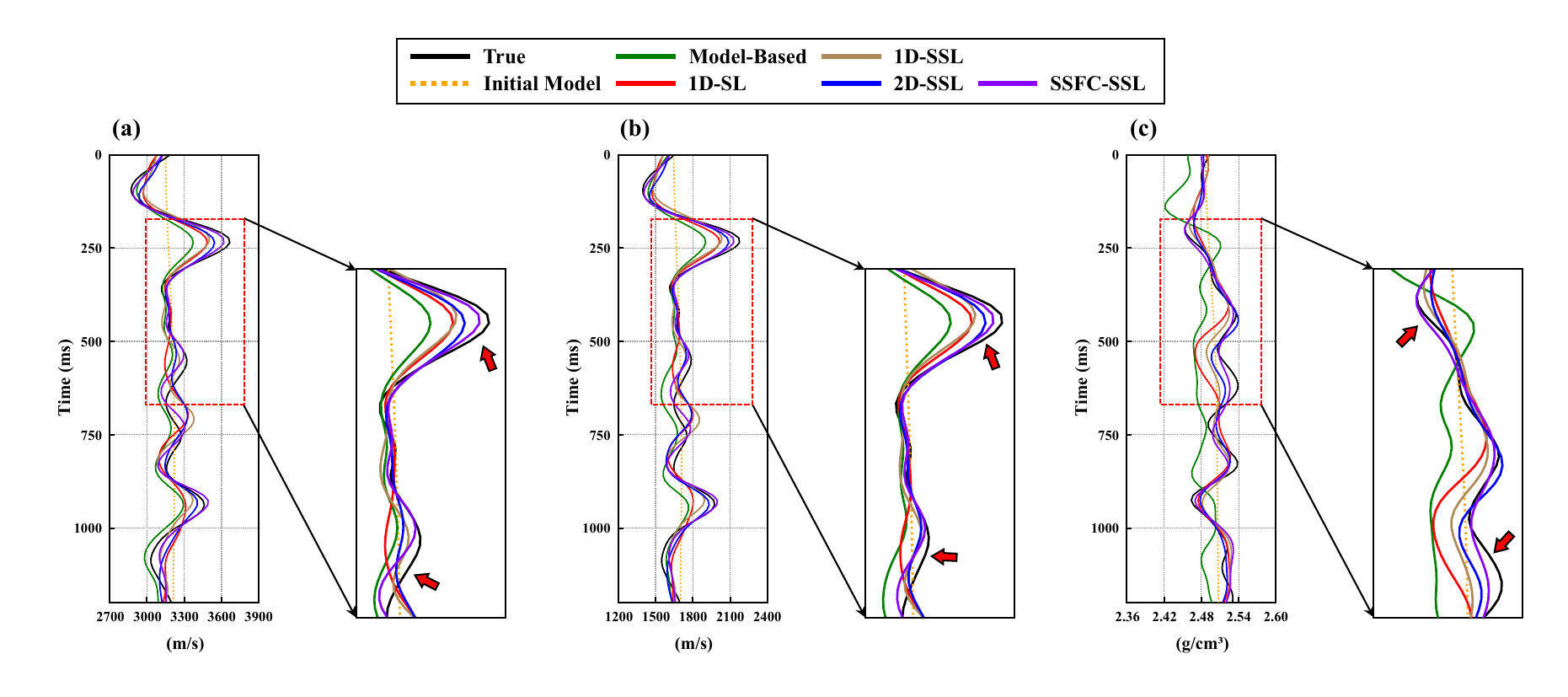}
\caption{Inversion results of (a) P-wave velocity, (b) S-wave velocity, and (c) density at well G2.}
\label{fig21}
\end{figure*}

\section{Experiments}
\subsection{Performance metrics and hyperparameter.}

\begin{table*}[!t]
\caption{Quantitative analysis of the inversion results for the field data.}
\renewcommand\arraystretch{1.2}
\centering
\label{table3}
\begin{tabular}{c|ccc|ccc|ccc}
\toprule[1.5pt]
\multirow{2}{*}{\textbf{Method}} & \multicolumn{3}{c|}
{\textbf{P-wave Velocity}}                                                 & \multicolumn{3}{c|}{\textbf{S-wave Velocity}}                                                 & \multicolumn{3}{c|}{\textbf{Density}}                                                         \\ \cline{2-10} 
                                 & \multicolumn{1}{c|}{PCC}             & \multicolumn{1}{c|}{R²}              & RMSE            & \multicolumn{1}{c|}{PCC}             & \multicolumn{1}{c|}{R²}              & RMSE            & \multicolumn{1}{c|}{PCC}             & \multicolumn{1}{c|}{R²}              & RMSE            \\ \hline
Model-Based                      & \multicolumn{1}{c|}{0.7979}          & \multicolumn{1}{c|}{0.7582}          & 0.0348          & \multicolumn{1}{c|}{0.7512}          & \multicolumn{1}{c|}{0.6948}          & 0.0672          & \multicolumn{1}{c|}{0.5176}          & \multicolumn{1}{c|}{0.4266}          & 0.0155          \\ \hline
1D-SL                            & \multicolumn{1}{c|}{0.8738}          & \multicolumn{1}{c|}{0.8203}          & 0.0297          & \multicolumn{1}{c|}{0.8767}          & \multicolumn{1}{c|}{0.7751}          & 0.0485          & \multicolumn{1}{c|}{0.8211}          & \multicolumn{1}{c|}{0.7401}          & 0.0073          \\ \hline
1D-SSL                           & \multicolumn{1}{c|}{0.9003}          & \multicolumn{1}{c|}{0.8496}          & 0.0253          & \multicolumn{1}{c|}{0.8868}          & \multicolumn{1}{c|}{0.7942}          & 0.0416          & \multicolumn{1}{c|}{0.8717}          & \multicolumn{1}{c|}{0.7650}          & 0.0049          \\ \hline
2D-SSL                           & \multicolumn{1}{c|}{0.9294}          & \multicolumn{1}{c|}{0.8652}          & 0.0205          & \multicolumn{1}{c|}{0.9103}          & \multicolumn{1}{c|}{0.8454}          & 0.0316          & \multicolumn{1}{c|}{0.9227}          & \multicolumn{1}{c|}{0.8516}          & 0.0037          \\ \hline
SSFC-SSL                         & \multicolumn{1}{c|}{\textbf{0.9478}} & \multicolumn{1}{c|}{\textbf{0.8981}} & \textbf{0.0130} & \multicolumn{1}{c|}{\textbf{0.9181}} & \multicolumn{1}{c|}{\textbf{0.8698}} & \textbf{0.0202} & \multicolumn{1}{c|}{\textbf{0.9450}} & \multicolumn{1}{c|}{\textbf{0.8838}} & \textbf{0.0031} \\ \bottomrule[1.5pt]
\end{tabular}
\end{table*}
\end{sloppypar}
In the experiment, to ensure the stable convergence of the network, we normalized the data using the following equations:
\begin{equation}\label{h19}
{m}^{\prime}= \frac{m - \mu_{m}}{\sigma_{m}}
\end{equation}
\begin{equation}\label{h20}
D^{\prime}= \frac{D - \mu_{D}}{\sigma_D}
\end{equation}
where $m$ and $D$ represent the original seismic data and the original model (or low-frequency model), $\mu_{m}$ and $\mu_{D}$ represent the mean of $m$ and $D$, and $\sigma_M$ and $\sigma_D$ represent the variance of $m$ and $D$. $m^{\prime}$ and $D^{\prime}$ denote the normalized model (or low-frequency model) and the seismic data. It is important to note that $m$ includes $V_P$, $V_S$, and $\rho$, which have different dimensions, so each parameter is normalized separately.

To quantify the accuracy of the inversion results, we introduced the Pearson correlation coefficient ($\mathrm{PCC}$), coefficient of determination ($\mathrm{R^{2}}$), and root mean square error ($\mathrm{RMSE}$). $\mathrm{PCC}$ is a measure of the linear relationship between two variables, quantifying the strength and direction of their correlation. It is defined as
\begin{equation}\label{H21}
\mathrm{PCC} = \frac{\sum_{k=1}^{n}(m_{k}- \mu _{m})(\hat{m_{k}} - \mu _{\hat{m}})}{\sqrt{\sum _{k=1}^{n}(y_{k}- \mu _{m})^{2}}\sqrt{\sum _{k=1}^{n}(\hat{m_{k}}- \mu _{\hat{m}})^{2}}}
\end{equation}
where $m_{k}$ is the $k$th true value, $\hat{m}_{k}$ represents the $k$th predicted value, $\mu _{m}$ and $\mu _{\hat{m}}$ are the mean values of the input parameters and predicted values, respectively, and $n$ denotes the number of data points. The $\mathrm{R^{2}}$ score is a statistical indicator used to measure the prediction effect of a regression model, representing the degree to which the model explains the total variation of the target variable. Its value ranges from 0 to 1, and the closer the $\mathrm{R^{2}}$ score is to 1, the closer the prediction results are to the true values. The definition of $\mathrm{R^{2}}$ is
\begin{equation}\label{h22}
\mathrm{R^{2}} = 1- \frac{\sum _{k=1}^{n}(m_{k}- \hat{m}_{k})^{2}}{\sum _{k=1}^{n}(m_{k} - \mu_{m})^{2}}
\end{equation}
The $\mathrm{RMSE}$ was obtained by taking the square root of the mean-square error (MSE), which is a measure of the difference between the predicted value and the true value. The $\mathrm{RMSE}$ can be formulated as
\begin{equation}\label{H23}
\mathrm{RMSE} = \sqrt{\frac{1}{n}\sum _{k=1}^{n}(m_{k}- \hat{m}_{k})^{2}}
\end{equation}

% Generally, $c_{1}$ is equal to 0.01, and $c_{2}$ is equal to 0.03 \cite{wang2004image}.

For network training, we set a series of hyperparameters, as listed in Table \ref{table1}. 
% The number of epochs for SSFC-SSL was set to 500.  The initial learning rates of the three sub-networks were all 0.002, the weight decay was 0.0001, and the dropout rate was 0.2. The batch size was 100,
Additionally, the number of multi-traces was set to 5. The experiment was conducted on a computer running the Windows 11 operating system, with a configuration that included 32GB of RAM, an Intel Core i5-12490F processor, and an NVIDIA GeForce GTX 3060 Graphics Processing Unit (GPU) with 8GB of video memory. The experimental platform used was PyTorch\cite{paszke2019pytorch}.

\subsection{Testing on Marmousi2 model}
To verify the effectiveness of the SSFC-SSL method, we used the Marmousi2 model data for inversion tests. This model includes 2720 common depth points (CDPs) and 600 sampling points, with a sampling interval of 2 ms. Reflection coefficients were calculated from the model using (\ref{H2}) and convolved with a zero-phase Ricker wavelet of 20 Hz dominant frequency. As shown in Fig. \ref{fig7}(a)-(e), five sets of seismic data with incident angles ranging from 0 to 24 degrees (at equal intervals) were synthesized as input. Additionally, a low-pass filter was applied to the model data to obtain the low-frequency model, which is shown in Fig. \ref{fig7}(f)-(h). As shown in Fig. \ref{fig8}, we extracted 18 traces at equal intervals to construct the labeled dataset, which accounted for only 0.66\% of the total traces. The remaining traces were randomly sampled at an interval of 50 traces as the unlabeled data. Fig. \ref{fig9} shows the loss curves of SSFC-SSL, and all losses can converge stably.

To assess the effectiveness of the proposed method, we compared it with one-dimensional supervised learning (1D-SL), one-dimensional semi-supervised learning (1D-SSL), and two-dimensional semi-supervised learning (2D-SSL) methods. Additionally, we used the commonly used model-driven BFGS method as a benchmark. Fig. \ref{fig10} shows the inversion results for different methods. Model-based methods can display the stratigraphic structure, but their resolution is slightly low. The 1D-SL method improves inversion resolution but brings noticeable vertical artifacts. Although the 1D-SSL method reduces these artifacts to some degree, it is still blurred at the rock layer unconformities, as shown by the elliptical wireframes in Fig. \ref{fig10}. The 2D-SSL and SSFC-SSL methods improve lateral continuity in the inversion results. Compared to the 2D-SSL method, the SSFC-SSL method offers clearer delineation at the boundaries of thin layers (red arrows, Fig. \ref{fig10}(n)-(0)). Fig. \ref{fig11} shows the residuals between the inversion results of each method and the true values. The SSFC-SSL method exhibits low errors in both the reservoir and complex geological structures (red and blue rectangular boxes). Fig. \ref{fig12} shows the prediction results for trace 600 not included in the training data. By comparing the inversion results with the measured 2D profile and 1D trace, it is evident that the SSFC-SSL method yields more accurate results (indicated by the red arrow). Despite the extreme complexity of the structural features and significant lithological changes at depths of 620 ms, the inversion results using the SSFC-SSL method maintain high accuracy and resolution. To quantitatively compare the accuracy of each method, we calculated the mean values of the inversion profiles using $\mathrm{PCC}$, $\mathrm{R^{2}}$, and $\mathrm{RMSE}$. The results are listed in Table \ref{table2}. The SSFC-SSL method achieved the best values in all indicators.

In practical applications, the number of available well-log labels is often limited due to complex geological conditions and high well-log costs. Therefore, we further designed a comparison of various methods under the condition of scarce labels. As shown in Fig. \ref{fig13}, we sampled seven and three training traces at equal intervals, which represent only 0.26\% and 0.11\% of the total number of traces, respectively. Figs. \ref{fig14} and \ref{fig15} show the inversion results using seven and three training traces, respectively. As the number of training traces decreases, the vertical artifacts of the 1D-SL and 1D-SSL methods become significantly more severe. Although the 2D-SSL method improves lateral continuity, it still exhibits numerous high-value errors in complex strata (see the areas marked by the elliptical box). In contrast, the SSFC-SSL method maintains high accuracy and resolution by fully considering the relationships between adjacent traces. Fig. \ref{fig16} shows the quantitative comparison of different inversion methods. By comparison, we can see that the SSFC-SSL method achieves higher $\mathrm{PCC}$ and $\mathrm{R^{2}}$ values, along with a low $\mathrm{RMSE}$. This verifies the efficiency and accuracy of the proposed method under limited well-log labels.

\subsection{Testing on field data}
To verify the practicality of the SSFC-SSL method, we applied it to field data from the South China Sea. The main rock types in the field data are sandstone and mudstone, which feature complex lithology, significant thickness variations, and strong heterogeneity. Therefore, high accuracy and resolution of velocity and density are essential. We selected a line consisting of 630 common depth points (CDPs), passing through well G1 (CDP = 98), well G2 (CDP = 304), well G3 (CDP = 443), and well G4 (CDP = 615). Wells G1, G2, and G3 were used for training, while well G3 was used to evaluate the inversion results. Before inversion, the seismic data were processed using noise suppression, surface-consistent deconvolution, amplitude recovery, and resampling to improve the signal-to-noise ratio and resolution. The low-frequency information of the different elastic parameters is obtained by layer interpolation, extrapolation, and smoothing of the corresponding well-log curves. The seismic data for different angles are shown in Fig. \ref{fig17}, and the low-frequency profiles of P-wave velocity, S-wave velocity, and density are shown in Fig. \ref{fig18}. The hyperparameters were the same as those used in the synthetic data experiments. 

We performed AVO inversion on field data using the model-based, 1D-SL, 1D-SSL, 2D-SSL, and SSFC-SSL methods, and the results are shown in Figs. \ref{fig18}, \ref{fig19}, and \ref{fig20}. The model-based method and conventional DL-based methods effectively uncover stratigraphic features. However, they have limitations in resolution, particularly in the accuracy of density inversion. In contrast, the elastic parameters obtained using the SSFC-SSL method exhibit higher resolution and better lateral continuity, as indicated by the elliptically labeled regions. To achieve a further comparison, the model-based method and the DL-based method results at the location of the blind well G2 are shown in Fig. \ref{fig21}. Table \ref{table3} shows the $\mathrm{PCC}$, $\mathrm{R^{2}}$, and $\mathrm{SSIM}$ values between the inversion results at the well G2 and the well-log curves. The inversion results of SSFC-SSL can fit the well-log curves well, especially in the area indicated by the arrows. 

\section{Conclusion}
In this article, we introduce the SSFC-SSL method, a novel approach to AVO inversion that addresses the limitations of traditional one-dimensional and two-dimensional DL methods. Unlike one-dimensional DL methods, which suffer from a lack of lateral continuity, and two-dimensional DL methods, which fail to fully utilize well-log data, SSFC-SSL overcomes these challenges by integrating a label annihilation operator and a strong spatial feature constraints (SSFC) network. This approach effectively constrains both well-log data and adjacent seismic traces, thereby improving both lateral continuity and inversion accuracy. Experimental results on both synthetic data and field data show that the SSFC-SSL method is superior to the model-based, one-dimensional DL, and two-dimensional DL methods in terms of lateral continuity and inversion accuracy. In future work, we plan to extend the SSFC-SSL method to post-stack inversion, aiming to further enhance the precision of seismic parameter estimation.

\bibliography{SSFC_SSL}
\bibliographystyle{IEEEtran}

\end{document}